\documentclass[letterpaper,journal]{IEEEtran}
\usepackage{amsmath}
\usepackage{amsfonts,amssymb}
\usepackage{amsthm}
\usepackage{array}
\usepackage{multirow}
\usepackage{rotating}
\usepackage{tabularray}
\usepackage{adjustbox}
\usepackage{textcomp}
\usepackage{stfloats}
\usepackage{graphicx}
\usepackage{textcomp}
\usepackage{caption,subcaption}
\usepackage{xcolor}
\usepackage{float}
\usepackage{cite}
\usepackage{url}
\usepackage{esint}
\usepackage{anyfontsize}

\usepackage{algorithmicx}
\usepackage{algorithm}
\usepackage{algpseudocode} 
\newtheorem{theorem}{Theorem}
\newtheorem{lemma}[theorem]{Lemma}
\usepackage{verbatim}
\usepackage{enumitem}
\newlist{Properties}{enumerate}{2}
\setlist[Properties]{label=Property \arabic*., font=\textit, itemindent=*,itemsep=1.5mm}
\usepackage{float}
\begin{document}
% \receiveddate{14 May, 2024}
% \reviseddate{XX Month, XXXX}
% \accepteddate{XX Month, XXXX}
% \publisheddate{XX Month, XXXX}
% \currentdate{14 May, 2024}
% \doiinfo{OJCOMS.2024.011100}
%%%%%%%%%%%%%%%%%%%%%%%%%%%%%%%%%
%%%%%%%%%%%%%%%%%%%%%%%%%%%%%%%%
\title{Cislunar Communication Performance and System Analysis with Uncharted Phenomena}
%%%%%%%%%%%%%%%%%%%%%%%%%%%%%%%%%
%%%%%%%%%%%%%%%%%%%%%%%%%%%%%%%%%
\author{S. Gecgel Cetin\IEEEauthorrefmark{1,2}, \IEEEmembership{Graduate Student Member, IEEE}, Á. Vazquez-Castro\IEEEauthorrefmark{3}, \IEEEmembership{Senior Member, IEEE},\\ and G. Karabulut Kurt\IEEEauthorrefmark{2}, \IEEEmembership{Senior Member, IEEE}}
% \affil{Istanbul Technical University, 34469 Istanbul, Turkey}
% \affil{Polytechnique Montréal, Montréal, QC H3T 1J4, Canada}
% \affil{Autonomous University of Barcelona and Institute of Space Studies of Catalonia (IEEC-UAB), 08193 Barcelona, Spain}
% \corresp{CORRESPONDING AUTHOR: S. Gecgel Cetin (e-mail: gecgel16@itu.edu.tr).}
%\authornote{This work was supported by the Natural Sciences and Engineering Research Council of Canada (NSERC) Discovery Grant program.}
%%__%%__%%__%%__%%%%__%%__%%__%%__%%%%__%%__%%__%%__%%
\maketitle
%%__%%__%%__%%__%%%%__%%__%%__%%__%%%%__%%__%%__%%__%%
\begin{abstract}
The Moon and its surrounding cislunar space have numerous unknowns, uncertainties, or partially charted phenomena that need to be investigated to determine the extent to which they affect cislunar communication. These include temperature fluctuations, spacecraft distance and velocity dynamics, surface roughness, and the diversity of propagation mechanisms. To develop robust and dynamically operative Cislunar space networks (CSNs), we need to analyze the communication system by incorporating inclusive models that account for the wide range of possible propagation environments and noise characteristics. In this paper, we consider that the communication signal can be subjected to both Gaussian and non-Gaussian noise, but also to different fading conditions. First, we analyze the communication link by showing the relationship between the brightness temperatures of the Moon and the equivalent noise temperature at the receiver of the Lunar Gateway. We propose to analyze the ergodic capacity and the outage probability, as they are essential metrics for the development of reliable communication. In particular, we model the noise with the additive symmetric alpha-stable distribution, which allows a generic analysis for Gaussian and non-Gaussian signal characteristics. Then, we present the closed-form bounds for the ergodic capacity and the outage probability. Finally, the results show the theoretically and operationally achievable performance bounds for the cislunar communication. To give insight into further designs, we also provide our results with comprehensive system settings that include mission objectives as well as orbital and system dynamics.
\end{abstract}
%%__%%__%%__%%__%%%%__%%__%%__%%__%%%%__%%__%%__%%__%%
%%%%%%%%%%%%%%%%%%%%%%%%%%%%%%%%%%%%%%%%%%%%%%%%%%%%%%
%%__%%__%%__%%__%%%%__%%__%%__%%__%%%%__%%__%%__%%__%%
%%%%%%%%%%%%%%%%%%%%%%%%%%%%%%%%%%%%%%%%%%%%%%%%%%%%%%%%%%%%%%%%%%%%%%%%
%%%%%%%%%%%%%%%%%%%%%%%%%%%%%%%%%%%%%%%%%%%%%%%%%%%%%%%%%%%%%%%%%%%%%%%%
%%%%%%%%%%%%%%%%%%%%%%%%%%%%%%%%%%%%%%%%%%%%%%%%%%%%%%%%%%%%%%%%%%%%%%%%
%%%%%%%%%%%%%%%%%%%%%%%%%%%%%%%%%%%%%%%%%%%%%%%%%%%%%%%%%%%%%%%%%%%%%%%%
%%%%%%%%%%%%%%%%%%%%%%%%%%%%%%%%%%%%%%%%%%%%%%%%%%%%%%%%%%%%%%%%%%%%%%%%
%%%%%%%%%%%%%%%%%%%%%%%%%%%%%%%%%%%%%%%%%%%%%%%%%%%%%%%%%%%%%%%%%%%%%%%%
%%%%%%%%%%%%%%%%%%%%%%%%%__Introduction__%%%%%%%%%%%%%%%%%%%%%%%%%%%%%%%
%%%%%%%%%%%%%%%%%%%%%%%%%%%%%%%%%%%%%%%%%%%%%%%%%%%%%%%%%%%%%%%%%%%%%%%%
%%%%%%%%%%%%%%%%%%%%%%%%%%%%%%%%%%%%%%%%%%%%%%%%%%%%%%%%%%%%%%%%%%%%%%%%
%%%%%%%%%%%%%%%%%%%%%%%%%%%%%%%%%%%%%%%%%%%%%%%%%%%%%%%%%%%%%%%%%%%%%%%% 
%%%%%%%%%%%%%%%%%%%%%%%%%%%%%%%%%%%%%%%%%%%%%%%%%%%%%%%%%%%%%%%%%%%%%%%%
%%%%%%%%%%%%%%%%%%%%%%%%%%%%%%%%%%%%%%%%%%%%%%%%%%%%%%%%%%%%%%%%%%%%%%%%
%%%%%%%%%%%%%%%%%%%%%%%%%%%%%%%%%%%%%%%%%%%%%%%%%%%%%%%%%%%%%%%%%%%%%%%%
\section{INTRODUCTION}\label{intro}
\IEEEPARstart{M}{oon} is coming to the fore with long-term and advanced goals, not least due to the involvement of new players on the commercial side \cite{Intro6, Intro7,sennur}. Lunar missions with economic and scientific objectives increase, but also emerge as precursors of deep space missions and expand the scope \cite{moon-mars, mars_nasa,yuan}. For example, the Artemis Base Camp, the surface habitation on the Moon, is being planned under the leadership of NASA and will also be used for missions to Mars \cite{2ndsub_2}. The catalysis of commercial and national actors accelerates a thriving progress to build cislunar space networks (CSNs) for potential users in space such as astronauts, crew exploration vehicles, robotic rovers and crewmember landers \cite{2ndsub_1}. CSNs provide communication, navigation and tracking services and are established with relay satellites, orbiters, spacecraft, terrain vehicles and rovers \cite{Intro3}. All architectures for CSNs are expected to ensure end-to-end security, cross-support as well as scalability and expandability to integrate them into future CSNs \cite{me1,baris1,Intro4}. However, the most urgent objectives for CSNs are not these, but dynamic capabilities \cite{ref1_t1} and high interoperability \cite{ref2_t1}, as they are the prerequisite to successfully conduct lunar missions.
%%%%%%%%%%%%%%%%%%%%%%%%%%%%%%%%%%%%%%%%%%%%%%%%%%%%%%%%%%%%%%%%%%%%%%%%
\subsection{Cislunar Space for CSNs}
CSNs must be designed and developed to withstand environmental conditions in cislunar space, but also to potential threats and anomalies such as solar scintillation \cite{solarscin1,solarscin2,solarburst1,solar2}, dust storms \cite{dust1}, natural and artificial radiation sources \cite{bib_5,bib_19_2,DSN,noisecharacter1}. Conveying the current technologies used for communications on Earth and in near space is the main approach to develop robust and adaptable systems with high performance \cite{Intro4,ref1_t1,ref2_t1}. However, previous missions show the differences between cislunar and near space, but also between the Earth and the Moon, which cannot be considered identical. Scientists and engineers design lunar communication systems by simulating the systems or by using appropriate models based on the accumulated knowledge. The tough challenge is to bring the interdisciplinary knowledge together for a realistic analysis. There are still partially charted phenomena, uncertainties and unknowns that can affect communication performance or jeopardize the entire mission \cite{unknowns1,unknowns2,unknowns3,unknowns4,uncertain1}.
%%%%%%%%%%%%%%%%%%%%%%%%%%%%%%%%%%%%%%%%%%%%%%%%%%%%%%%%%%%%%%%%%%%%%%%%
\begin{figure*}[!ht]
\centering
\includegraphics[width=\linewidth]{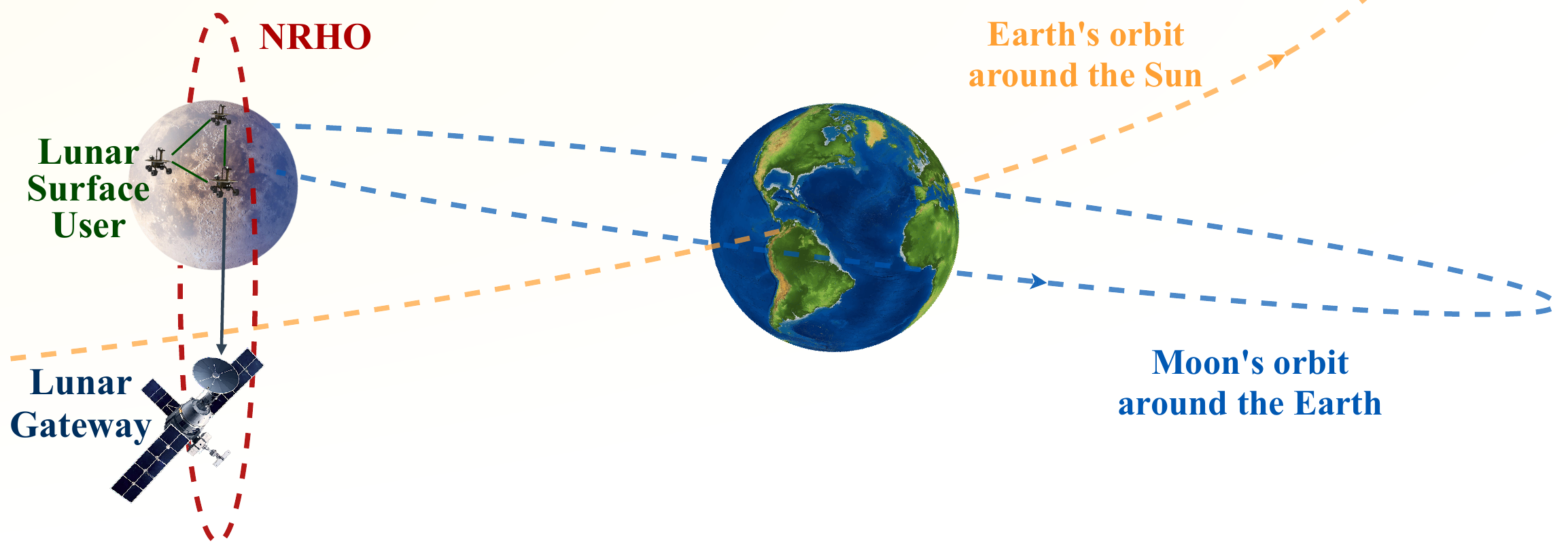}
\caption{The uplink communication from the lunar surface user to the Lunar Gateway. The exaggerated view of cislunar space in which the Lunar Gateway orbits in near-rectilinear halo orbit (NRHO) while the Moon and Earth move in their orbits. One of the lunar surface users represents the communication terminal of the lunar mission and transmits information.}
\label{SYSTEM}
\end{figure*}
%%%%%%%%%%%%%%%%%%%%%%%%%%%%%%%%%%%%%%%%%%%%%%%%%%%%%%%%%%%%%%%%%%%%%%%%

The unique conditions of cislunar space bring with them crucial aspects that must be considered in the development of the communication system to ensure robust and reliable communication. The physical temperatures of the Moon and Earth change daily and seasonally as they orbit around the Earth and Sun, respectively \cite{tempfluct2,orbiting1,surface1}. However, the lack of a dense atmosphere and large bodies of water as on Earth leads to strong temperature fluctuations on the Moon during the lunar day and night \cite{tempfluct1}. This inevitably affects the operation of electronic systems, even those produced with large survivability limits \cite{bib_1,thermal_elect2,thermal_elect3}. Such temperature fluctuations also pose a major challenge to the stable operation of communication systems, as the noise signals vary depending on the system temperature \cite{Tb1,R7_Morabito,R8_Morabito,Intro9}. This is a stark contrast with communication systems on Earth and in near space, where environmental conditions are more stable and predictable. Furthermore, due to the hilly terrain and the spherical shape of the Moon, the lunar surface is not evenly exposed to sunlight. As a result, the temperature fluctuations between sunlit and shadowed areas are complicated during the lunar day and night \cite{bib_20surface2}. For example, the system temperature can change instantly and cause impulsive noises when a rover moves to another location, even at close range. This is another difference to communications on Earth and in near space, where the noise is usually assumed to be Gaussian. To provide reliable and robust communication given the dynamic nature of the cislunar environment, we should consider the presence of both non-Gaussian and Gaussian signal characteristics.

Another important aspect is the diversity of propagation mechanisms in the cislunar environment. For example, the visibility of transmission paths is uncertain due to the hilly terrain and the mobility of spacecraft and users on the surface \cite{baris3,Multipath1,multipath2,RiceK4}. Thus, line-of-sight (LoS) and non-line-of-sight (NLoS) transmissions can occur individually or together. %\cite{RiceK1,RiceK2,RiceK3,RiceK4,bib_21,bib_22}. 
We also know that diffraction loss increases with decreasing elevation angle \cite{RiceK4}. In addition, the multifaceted lunar surface and the varying reflectivity of the lunar regolith further complicate the scattering and reflection conditions \cite{RiceK1,RiceK2,RiceK3,RiceK4,bib_21,bib_22,bib_3}. However, communication systems must be developed to ensure the expected performance even under unfavorable conditions for the signal attenuation. Therefore, we should consider (at least the most probable cases) for LoS and NLoS communication with other attenuation effects such as scattering, diffraction and reflection properties.
%%%%%%%%%%%%%%%%%%%%%%%%%%%%%%%%%%%%%%%%%%%%%%%%%%%%%%%%%%%%%%%%%%%%%%%%
\subsection{Related Works and System Considerations}\label{related}
Aforementioned aspects of cislunar space and the corresponding suggestions emphasize the need for inclusive models that account for a wide range of propagation conditions and noise characteristics. In this way, we can design and develop robust and reliable cislunar communication systems but also dynamically adaptable to support the short- and long-term goals of lunar missions and beyond \cite{intro1}.

Communication systems in near space and on Earth are generally developed under the assumption of Gaussian noise. However, this may not apply to all communication scenarios, resulting in poor performance. In particular, communication in challenging environments such as cislunar space with temperature fluctuations, unstable orbital conditions \cite{unstable1_nrho,unstable2_nrho}, unexpected interference and anomalies \cite{RFI1,RFI2,bib_6,bib_19,Intro10} that can lead to significant deviations in noise characteristics. Symmetric Alpha-Stable ($\text{S}\alpha\text{S}$) distribution with its rich parameterization proves to be a well-suited model for relatively unstable systems such as power-line, underwater and cislunar communications. It enables generic analysis and adaptive development of communication systems that can operate effectively under different noise conditions, from common Gaussian noise scenarios to more extreme cases with impulsive, non-Gaussian noise. Therefore, we model the channel with additive symmetric alpha-stable noise ($\text{AS}\alpha\text{SN}$) to consider both Gaussian and non-Gaussian signals.

$\text{AS}\alpha\text{SN}$ is usually adopted to model interference, mostly within the additive white Gaussian noise (AWGN) channel and the studies provide analyses over different performance metrics such as error rates, outage, or detection probabilities \cite{bib_10,bib_11,bib_12,bib_14,refgrup1_1,refgrup1_2}. There are some studies that perform analysis on $\text{AS}\alpha\text{SN}$ channels without Gaussian noise. In \cite{refgrup2_3}, the authors analyze the diversity combining schemes under Rayleigh fading and $\text{AS}\alpha\text{SN}$ and present the error performance with closed form expressions for the diversity and array gains. In \cite{refgrup2_2}, the error performance of phase shift keying modulation is investigated using the geometric power of $\text{AS}\alpha\text{SN}$ variables without considering fading. However, the majority of studies focus on error performance rather than capacity or ergodic capacity. There are only a handful of studies that consider $\text{AS}\alpha\text{SN}$ within the basis of AWGN channel \cite{CAP3,CAP1,CAP2,CAP4}, due to challenging properties of $\text{S}\alpha\text{S}$ variables. As far as we are aware, \cite{Capacity_bound} is the first study that directly focuses on the optimization of the capacity maximization problem without system constraints or other focuses. The authors of \cite{Capacity_bound} use the generalization of Shannon’s channel coding theorem for non-Gaussian channels \cite{TeSun} and adapt this generalization for $\text{AS}\alpha\text{SN}$. In this study, we use this bound as the basis for the theoretical analyzes under the Nakagami-m fading.

In the existing studies on lunar communication, the Rician fading model is usually considered, while Rayleigh fading is assumed for cases where there is no LoS, such as at low elevation angles \cite{Multipath1,multipath2,RiceK4,RiceK1,RiceK2,RiceK3,bib_21,bib_22,bib_3}. However, all these studies focus on the direct communication link between the Earth station and the Lunar South Pole. Since the Earth and the Moon rotate at approximately the same speed, the relative movements of the transmitter and receiver do not change or are negligible. For CSNs providing services for complex missions, the situation is quite different. In particular, CSN architectures are based on relay spacecrafts moving at varying speeds in unstable lunar orbits. If we consider the cislunar spacecraft and its relative movement to the mobile user on the lunar surface, the propagation mechanisms become even more diverse. The Nakagami-m fading model is feasible to exactly or approximately represent different propagation scenarios as well as for closed-form theoretical analyses. Therefore, we performed our analyzes with the Nakagami-m fading model to gain a better insight into the performance range of cislunar communication. \textit{To the best of our knowledge, there is no study that specifically addresses capacity analysis under $\text{AS}\alpha\text{SN}$ and Nakagami-m fading, nor is there any study that provides an analysis of ergodic capacity and outage probability for cislunar communication with this extent.}
%%%%%%%%%%%%%%%%%%%%%%%%%%%%%%%%%%%%%%%%%%%%%%%%%%%%%%%%%%%%%%%%%%%%%%%%
\subsection{Contributions}
We perform all analyzes via NASA's Lunar Gateway communication link, as illustrated in Fig. \ref{SYSTEM}. Our main contributions are listed below.
\begin{itemize}
\item We highlight the critical system aspects for the design and development of CSNs but also provide insights into the potential threats and anomalies in cislunar space.
\item We analyze the communication link of the foremost CSN architecture by presenting the theoretical formulation for the relationship between the brightness temperatures and the equivalent noise temperature at the receiver. In this way, we address a less unexplored aspect of communication in cislunar space.
\item We provide the theoretical derivations in closed-form for the probability distribution function (pdf) of the instantaneous signal-to-noise ratio (SNR), the lower bound of the ergodic capacity and the corresponding upper bound of the outage probability for $\text{AS}\alpha\text{SN}$ and Nakagami-m fading, which to the best of our knowledge are presented for the first time in the literature.
\item As described in Algorithm 1, we present the numerical approximation algorithm for the ergodic capacity to account for fading by adapting the Blahut-Arimato algorithm, which to our knowledge is the first in the literature.
\item We establish the direct link between environmental variations (temperature, noise and fading) in cislunar space and the corresponding performance metrics via our results.
\item We extend the our results to provide realistic insights for system-level designs, which affects the choice of technologies, frequency band and power levels, as well as for mission-level considerations.
\end{itemize}
Our analyzes show the obvious influence of brightness temperature under $\text{AS}\alpha\text{SN}$ for the S and Ka bands, which are intended for low and high data rates. The ergodic capacity decreases and the outage probability increases significantly depending on the brightness temperature for the Gaussian and non-Gaussian noise characteristics. Furthermore, our results show how much fading conditions amplify these effects under different system settings. Overall, our work provides insight into the design and optimization of future cislunar communication systems to meet the upward trend of expected performance.
%%%%%%%%%%%%%%%%%%%%%%%%%%%%%%%%%%%%%%%%%%%%%%%%%%%%%%%%%%%%%%%%%%%%%%%%
\subsection{Notation and Organization}
The rest of the article is structured as follows. The preliminaries are given in Section \ref{prelim}. Section \ref{system_and_channel} presents an unified modeling framework that generalizes system and channel models by considering the conditions in cislunar space and the effect of lunar illumination. Section \ref{perform} provides the theoretical analysis for the ergodic capacity and outage probability together with the numerical approximations for ergodic capacity. The system-level performance analysis is performed by integrating the theoretical analysis and the numerical results are presented in Section \ref{results}. Finally, Section \ref{conclusion} concludes the paper.

\textit{Notation:}
Throughout the paper, absolute value of a scalar is denoted by $\left |{ \cdot }\right |$ and $\mathbb{E}\left [ X \right]$, $F_X(x)$ and $f_X(x)$ denote the mean, the cumulative distribution function (cdf) and pdf of a random variable $X$. $\Gamma (\cdot)$, $\Gamma (\cdot,\cdot)$ and $G_{p,q}^{m,n} \left[ (\cdot) \left| (\cdot) \right.\right] $ symbolize the gamma function, the lower incomplete gamma function and the Meijer-G function. 
%%%%%%%%%%%%%%%%%%%%%%%%%%%%%%%%%%%%%%%%%%%%%%%%%%%%%%%%%%%%%%%%%%%%%%%%
%%%%%%%%%%%%%%%%%%%%%%%%%%%%%%%%%%%%%%%%%%%%%%%%%%%%%%%%%%%%%%%%%%%%%%%%
%%%%%%%%%%%%%%%%%%%%%%%%%%%%%%%%%%%%%%%%%%%%%%%%%%%%%%%%%%%%%%%%%%%%%%%%
%%%%%%%%%%%%%%%%%%%%%%%%%%%%%%%%%%%%%%%%%%%%%%%%%%%%%%%%%%%%%%%%%%%%%%%%
%%%%%%%%%%%%%%%%%%%%%%%%%%%%%%%%%%%%%%%%%%%%%%%%%%%%%%%%%%%%%%%%%%%%%%%%
%%%%%%%%%%%%__Problem formulation and Background__%%%%%%%%%%%%%%%%%%%%%%
%%%%%%%%%%%%%%%%%%%%%%%%%%%%%%%%%%%%%%%%%%%%%%%%%%%%%%%%%%%%%%%%%%%%%%%%
%%%%%%%%%%%%%%%%%%%%%%%%%%%%%%%%%%%%%%%%%%%%%%%%%%%%%%%%%%%%%%%%%%%%%%%%
%%%%%%%%%%%%%%%%%%%%%%%%%%%%%%%%%%%%%%%%%%%%%%%%%%%%%%%%%%%%%%%%%%%%%%%%
%%%%%%%%%%%%%%%%%%%%%%%%%%%%%%%%%%%%%%%%%%%%%%%%%%%%%%%%%%%%%%%%%%%%%%%%
%%%%%%%%%%%%%%%%%%%%%%%%%%%%%%%%%%%%%%%%%%%%%%%%%%%%%%%%%%%%%%%%%%%%%%%%
%%%%%%%%%%%%%%%%%%%%%%%%%%%%%%%%%%%%%%%%%%%%%%%%%%%%%%%%%%%%%%%%%%%%%%%%
\section{PRELIMINARIES}
\label{prelim}
The alpha-stable distribution, is denoted by ${\cal S}(\alpha, \beta, \lambda, \mu)$ where the index of the stability $\alpha \in (0,2]$, the skewness parameter $\beta \in[-1, +1]$, the scale parameter $\lambda \in(0, +\infty)$ and the shift parameter $\mu \in(-\infty, +\infty)$ are parameters. When $\alpha\neq 1$, the alpha-stable distribution is defined with the characteristic function as follows \cite{stable1}
\begin{equation}
\Phi(t )= \exp \left [j\mu t-\lambda^{\alpha}\vert t \vert ^{\alpha} \left ( 1-j\beta \rm{sign}(t) \tan\frac{\pi \alpha}{2} \right ) \right ].
\end{equation}
If $\beta=0$ and $\mu=0$, the alpha-stable distribution is called as symmetric alpha-stable ($\text{S}\alpha\text{S}$) distribution, which is a generalization of Gauss distribution with zero mean and is valid in many scenarios considering signal distortions. The characteristic function of $\text{S}\alpha\text{S}$ distribution takes the form as follows
\begin{equation}
\Phi(t) = \exp(-\lambda^{\alpha} \vert t \vert ^{\alpha}).
\end{equation}
Here, we provide five properties of alpha-stable random variables that are used throughout the paper.
%%__%%__%%__%%__%%
%%__%%__%%__%%__%% 
%%__%%__%%__%%__%%
\begin{Properties}
\item Let $Z \sim S(\alpha, \beta, \lambda, \mu)$ with $\alpha \in (0,2)$. Then,
$\begin{matrix}
& & & {E}[|Z|^p] <\infty & \text{ for any} & 0<p<\alpha,
\\ 
& & & {E}[|Z|^p]= \infty & \text{ for any} & p\geq \alpha.
\end{matrix} $ 
%%%%%%%%%%%%%%%%%%%%%%%%%%%%%%%%%%%%%%%%%%%%%%
\item Let $Z \sim S(\alpha, \beta, \lambda, \mu)$ with $\alpha \in (0,2)$. Then, the shift parameter $\mu$ equals to the mean.
%%%%%%%%%%%%%%%%%%%%%%%%%%%%%%%%%%%%%%%%%%%%%%
\item Let $Z \sim S(\alpha ,0,\lambda,0)$ with $\alpha \in (1,2]$. Then,
\begin{equation} 
\mathbb {E}[|Z|] = \frac {2 \lambda \Gamma \left ({1 - \frac {1}{\alpha }}\right )}{\pi }. 
\end{equation}
%%%%%%%%%%%%%%%%%%%%%%%%%%%%%%%%%%%%%%%%%%%%%% 
\item Let $Z_1$ and $Z_2$ independent and identically distributed \text{(i.i.d.)} variables of $Z_i \sim S(\alpha, \beta_i, \lambda_i, \mu_i)$ with $\alpha \in (1,2]$ for $i=1,2$. Then, $Z_1 + Z_2 \sim S(\alpha,\beta,\lambda,\mu)$ where $\beta = \frac{\beta_1\lambda_1^\alpha + \beta_2\lambda_2^\alpha}{\lambda_1^\alpha + \lambda_2^\alpha}$, $\lambda = (\lambda_1^\alpha + \lambda_2^\alpha)^{1/\alpha}$, $\mu=\mu_1+\mu_2$. 
%%%%%%%%%%%%%%%%%%%%%%%%%%%%%%%%%%%%%%%%%%%%%% 
\item The pdf of a $\text{S}\alpha\text{S}$ variable, $Z \sim S(\alpha ,0,\lambda,0)$ with $\alpha \in (1,2]$ is defined as
\begin{equation}
 p_{Z}(z)=\frac {1}{2\pi }\int _{-\infty }^{\infty }e^{-|\lambda t|^{\alpha }}e^{-jtz}dt.
\end{equation}
\end{Properties}
%%__%%__%%__%%__%%
%%__%%__%%__%%__%% 
%%__%%__%%__%%__%%
%%%%%%%%%%%%%%%%%%%%%%%%%%%%%%%%%%%%%%%%%%%%%%%%%%%%%%%%%%%%%%%%%%%%%%%%
%%%%%%%%%%%%%%%%%%%%%%%%%%%%%%%%%%%%%%%%%%%%%%%%%%%%%%%%%%%%%%%%%%%%%%%%
%%%%%%%%%%%%%%%%%%%%%%%%%%%%%%%%%%%%%%%%%%%%%%%%%%%%%%%%%%%%%%%%%%%%%%%%
%%%%%%%%%%%%%%%%%%%%%%%%%%%%%%%%%%%%%%%%%%%%%%%%%%%%%%%%%%%%%%%%%%%%%%%%
%%%%%%%%%%%%%%%%%%%%%%%%%%%%%%%%%%%%%%%%%%%%%%%%%%%%%%%%%%%%%%%%%%%%%%%%
%%%%%%%%%%%%%%%%%%%%%%%%%%%%%%%%%%%%%%%%%%%%%%%%%%%%%%%%%%%%%%%%%%%%%%%%
%%%%%%%%%%%%%%%%%%%__System Model and Analysis__%%%%%%%%%%%%%%%%%%%%%%%%
%%%%%%%%%%%%%%%%%%%%%%%%%%%%%%%%%%%%%%%%%%%%%%%%%%%%%%%%%%%%%%%%%%%%%%%%
%%%%%%%%%%%%%%%%%%%%%%%%%%%%%%%%%%%%%%%%%%%%%%%%%%%%%%%%%%%%%%%%%%%%%%%%
%%%%%%%%%%%%%%%%%%%%%%%%%%%%%%%%%%%%%%%%%%%%%%%%%%%%%%%%%%%%%%%%%%%%%%%%
%%%%%%%%%%%%%%%%%%%%%%%%%%%%%%%%%%%%%%%%%%%%%%%%%%%%%%%%%%%%%%%%%%%%%%%%
%%%%%%%%%%%%%%%%%%%%%%%%%%%%%%%%%%%%%%%%%%%%%%%%%%%%%%%%%%%%%%%%%%%%%%%%
%%%%%%%%%%%%%%%%%%%%%%%%%%%%%%%%%%%%%%%%%%%%%%%%%%%%%%%%%%%%%%%%%%%%%%%%
\section{UNIFIED MODELING FRAMEWORK}\label{system_and_channel}
In this section, we provide an unified modeling framework and analyze the communication link shown in Fig. \ref{SYSTEM}, in which the receiver antenna of the Lunar Gateway is directed to a transmitter antenna of the user on the lunar surface. First, we introduce the generalized system model with the relationship between the lunar surface temperature and the receiver noise temperature. Considering this relationship as well as other potential threats and anomalies, we then present the generalized channel model for cislunar communication.
%%%%%%%%%%%%%%%%%%%%%%%%%%%%%%%%%%%%%%%%%%%%%%%%%%%%%%%%%%%%%%%%%%%%%%%%
%%%%%%%%%%%%%%%%%%%%%%%%%%%%%%%%%%%%%%%%%%%%%%%%%%%%%%%%%%%%%%%%%%%%%%%%
\subsection{Generalized System Model}\label{system_mod}
The power of the received signal at the Lunar Gateway is calculated as follows according to the Friis equation, taking the line of sight into account.
\begin{equation}
P_r = P_t \frac{G_t G_r c^2}{(4\pi f d)^{2} L_{t} L_{r}},
\end{equation}
where $c$ is the speed of light, $f$ is the frequency, $d$ is the distance, $L_{t}$ and $L_{r}$ represent the losses due to transmitter and receiver equipment. ${P_r}$ and $P_t$ denotes the power of the received and transmitted signal. {$G_t$ and $G_r$ are the antenna gain of transmitter and receiver that is given as
\begin{equation}
G = \eta_{A} \left ( \frac{D \pi f }{c} \right )^2 ,
\end{equation}
where $D$ is the diameter of the antenna and $\eta_{A}$ is the antenna aperture efficiency.}

The power of the noise for per Hertz is calculated according to the Rayleigh–Jeans law \cite{kt1,kt2} as
\begin{equation}
N_0 = k T_{op}, 
\end{equation}
where $k$ is the Boltzmann constant $(\approx 1.38 \times 10^{-23} J/K)$, and $T_{op}$ is the operational equivalent noise temperature. $T_{op}$ is calculated depending on cosmic microwave background noise temperature ($T_{CMB}$), antenna noise temperature ($T_{A}$), transmission line temperature ($ T_{TL}$), and receiver noise temperature ($T_R$) as follows 
\begin{equation}
%T_{op} \left (T_{A}\right) 
T_{op} = T_{CMB} + T_{A} + \frac{1}{\eta_{rad}}T_{TL} + \frac{1}{\eta_{rad} \eta_{TL}}T_R ,
\end{equation}
where $\eta_{rad}$ and $\eta_{TL}$ show the radiation efficiency and the thermal efficiency of the transmission line. 

When the receiver antenna has losses, antenna noise temperature $T_{A}$ counts external and internal noise temperatures in it. The external noise temperature of the antenna, $T_{A,external}$ is aroused from the brightness temperature\footnote[1]{The physical temperature results in the brightness temperature, depending on the radiation properties of the surface and the wavelength \cite{bib_3}.} of the subtended body $T_{B}(\theta_A,\varphi_A)$ and calculated as \cite{R1,R2_kraus,R3_ITU}
\begin{equation}
\label{Taexternal}
T_{A,external} = \frac{1}{\Omega_{A}} \underset{4\pi}{\oiint} \overline{F}(\theta_A,\varphi_A) T_{B}(\theta_A,\varphi_A) d\Omega.
\end{equation}
Here, $\Omega_{A}$ is the antenna solid angle and $\overline{F}(\theta_A,\varphi_A)$ is the antenna normalized power pattern depending on the observation angles $(\theta,\varphi)$.

If there are no unexpected sources of radiation, it can be assumed that $T_{B}(\theta_A,\varphi_A) = T_{B}$, the brightness temperature is the same in all directions of antenna surrounding \cite{R2_kraus}. Then, the equation (\ref{Taexternal}) is simplified and the increase in the external antenna noise temperature ($\Delta T_{A,external}$) is calculated as
\begin{equation}\label{Ta_ex_1}
\Delta T_{A,external} = 
\begin{cases}
T_{B}, & \Omega_M \gg \Omega_A \\ 
\frac{\Omega_B}{\Omega_A} T_{B}, & \text{ otherwise} 
\end{cases}
\end{equation}
where $\Omega_M$ and $\Omega_A$ represent the solid angles of the Moon and the antenna, respectively. However, the noise sources in radio astronomy show random polarization and this cause to receive half of power by the antenna. Therefore the correction factor is added as seen below \cite{R2_kraus}. 
\begin{equation}\label{Ta_ex_2}
\Delta T_{A,external} = 
\begin{cases}
\frac{1}{2} T_{B}, & \Omega_M \gg \Omega_A \\ 
\frac{\Omega_M}{2\Omega_A} T_{B}, & \text{ otherwise} 
\end{cases}
\end{equation}
$\Omega_A$ can be approximated in stredian depending on the half power beam width (HPBW) of the antenna in the planes of the observation angles as \cite{R2_kraus}
\begin{equation}\label{kraus}
\Omega_A \simeq \theta_{A,HPBW} \varphi_{A,HPBW}.
\end{equation}
The solid angle subtended by Moon at the distance $d_M$ for any $d_M\geq R_M$ is calculated in stredian as
\begin{equation}
\Omega_M = 2 \pi \left ( 1 - \frac{\sqrt{d_M^2 - R_M^2 }}{d_M^2} \right ),
\end{equation}
where $R_{M}$ is the radius of the Moon ($\approx 1737$ km). The internal antenna noise temperature, $T_{A,internal}$ is caused by the physical temperature of the antenna $T_{AP}$ and calculated as
\begin{equation}
T_{A,internal} = T_{AP}\left ( \frac{1}{\eta_{rad} } - 1 \right ).
\end{equation}
Here, $\eta_{rad}$ is radiation efficiency of the antenna. The transmission line noise temperature depends on the physical temperature of the transmission line ($T_{TLP}$) and is calculated as follows
\begin{equation}
T_{TL} = T_{TLP}\left ( \frac{1}{\eta_{TL} } - 1 \right ).
\end{equation}
As can be seen from the above equations, the noise signal is influenced by temperature fluctuations in the surrounding area of the transmitter to which the receiver antenna is directed. We can also heuristically notice that the noise shows impulsive characteristic, especially when the transmitter moves between sunlit and shadowed areas on the lunar surface.
%%%%%%%%%%%%%%%%%%%%%%%%%%%%%%%%%%%%%%%%%%%%%%%%%%%%%%%%%%%%%%%%%%%%%%%%
%%%%%%%%%%%%%%%%%%%%%%%%%%%%%%%%%%%%%%%%%%%%%%%%%%%%%%%%%%%%%%%%%%%%%%%%
%%%%%%%%%%%%%%%%%%%%%%%%%%%%%%%%%%%%%%%%%%%%%%%%%%%%%%%%%%%%%%%%%%%%%%%%
%%%%%%%%%%%%%%%%%%%%%%%%%%%%%%%%%%%%%%%%%%%%%%%%%%%%%%%%%%%%%%%%%%%%%%%%
%%%%%%%%%%%%%%%%%%%%%%%%%%%%%%%%%%%%%%%%%%%%%%%%%%%%%%%%%%%%%%%%%%%%%%%%
%%%%%%%%%%%%%%%%%%%%%%%%%%%%%%%%%%%%%%%%%%%%%%%%%%%%%%%%%%%%%%%%%%%%%%%%
%%%%%%%%%%%%%%%%%%%__Signal and Channel Model__%%%%%%%%%%%%%%%%%%%%%%%%%
%%%%%%%%%%%%%%%%%%%%%%%%%%%%%%%%%%%%%%%%%%%%%%%%%%%%%%%%%%%%%%%%%%%%%%%%
%%%%%%%%%%%%%%%%%%%%%%%%%%%%%%%%%%%%%%%%%%%%%%%%%%%%%%%%%%%%%%%%%%%%%%%%
%%%%%%%%%%%%%%%%%%%%%%%%%%%%%%%%%%%%%%%%%%%%%%%%%%%%%%%%%%%%%%%%%%%%%%%%
%%%%%%%%%%%%%%%%%%%%%%%%%%%%%%%%%%%%%%%%%%%%%%%%%%%%%%%%%%%%%%%%%%%%%%%%
%%%%%%%%%%%%%%%%%%%%%%%%%%%%%%%%%%%%%%%%%%%%%%%%%%%%%%%%%%%%%%%%%%%%%%%%
%%%%%%%%%%%%%%%%%%%%%%%%%%%%%%%%%%%%%%%%%%%%%%%%%%%%%%%%%%%%%%%%%%%%%%%%
\subsection{Generalized Channel Model}\label{channel_mod}
The presence of non-Gaussian and Gaussian signals should be considered for the aspects of communication in cislunar space. Therefore, we utilize generalization capability of the $\text{S}\alpha\text{S}$ distribution for non-Gaussian and Gaussian processes \cite{CAP3} and define the $\text{AS}\alpha\text{SN}$ variable as
\begin{equation}
n = \sqrt{A_{1}} G_{1} + j \sqrt{A_{2}} G_{2}.
\end{equation}
Here, $A_{1}$ and $A_{2}$ are i.i.d. alpha-stable distributed variables follow $S(\alpha /2,1,[\cos (\pi \alpha /4)]^{2/\alpha },0)$. $G_{1}$ and $G_{2}$ are i.i.d. Gaussian random variables follow $\mathcal {N}(0,\sigma ^{2})$. As proven in the Appendix A, $n$ follows $n \sim S \left ( \alpha ,0,2^{\left ( \frac{1}{\alpha}-\frac{1}{2} \right )}\sigma,0 \right )$. Note that for the special case of $\alpha = 2$, $n$ can also be defined as a Gaussian variable and represented as $n \sim \mathcal {N}(0,2\sigma ^{2})$ or $n \sim S(2,0,\sigma,0)$.
%%%%%%_______Appendix A_______%%%%%%%%
%%__%%__%%__%%__%%%%__%%__%%__%%__%%%%
%%%%%%%%%%%%%%%%%%%%%%%%%%%%%%%%%%%%%%
\renewcommand\qedsymbol{$\blacksquare$}
\begin{proof}
The proof is presented in Appendix A.
\end{proof}

The received signal at the cislunar spacecraft (Lunar Gateway) as follows
\begin{equation} 
 y = \sqrt{P_r}hx + n,
\end{equation}
where $x$ represents the modulated symbol and $h$ is the complex fading coefficient. We use Nakagami-m fading model, since it corresponds to the Rayleigh\footnote[2]{Nakagami-m fading model corresponds Rayleigh fading for $m = 1$.} fading but also can be approximated with Rician\footnote[3]{Nakagami-m fading model approximates Rician fading well for $m = {(K+1)^2}/{(2K+1)}$ where $K$ is the Rician fading parameter.}, mild or no\footnote[4]{Nakagami-m fading model converges to non-fading when $m \to \infty$.} fading cases \cite{alouini}.
%%%%%%%%%%%%%%%%%%%%%%%%%%%%%%%%%%%%%%%%%%%%%%%%%%%%%%%%%%%%%%%%%%%%%%%%
%%%%%%%%%%%%%%%%%%%%%%%%%%%%%%%%%%%%%%%%%%%%%%%%%%%%%%%%%%%%%%%%%%%%%%%%
%%%%%%%%%%%%%%%%%%%%%%%%%%%%%%%%%%%%%%%%%%%%%%%%%%%%%%%%%%%%%%%%%%%%%%%%
%%%%%%%%%%%%%%%%%%%%%%%%%%%%%%%%%%%%%%%%%%%%%%%%%%%%%%%%%%%%%%%%%%%%%%%%
%%%%%%%%%%%%%%%%%%%%%%%%%%%%%%%%%%%%%%%%%%%%%%%%%%%%%%%%%%%%%%%%%%%%%%%%
%%%%%%%%%%%%%%%%%%%%%%%%%%%%%%%%%%%%%%%%%%%%%%%%%%%%%%%%%%%%%%%%%%%%%%%%
%%%%%%%%%%%%%%%%%%%%%__Performance Analysis__%%%%%%%%%%%%%%%%%%%%%%%%%%%
%%%%%%%%%%%%%%%%%%%%%%%%%%%%%%%%%%%%%%%%%%%%%%%%%%%%%%%%%%%%%%%%%%%%%%%%
%%%%%%%%%%%%%%%%%%%%%%%%%%%%%%%%%%%%%%%%%%%%%%%%%%%%%%%%%%%%%%%%%%%%%%%%
%%%%%%%%%%%%%%%%%%%%%%%%%%%%%%%%%%%%%%%%%%%%%%%%%%%%%%%%%%%%%%%%%%%%%%%%
%%%%%%%%%%%%%%%%%%%%%%%%%%%%%%%%%%%%%%%%%%%%%%%%%%%%%%%%%%%%%%%%%%%%%%%%
%%%%%%%%%%%%%%%%%%%%%%%%%%%%%%%%%%%%%%%%%%%%%%%%%%%%%%%%%%%%%%%%%%%%%%%%
%%%%%%%%%%%%%%%%%%%%%%%%%%%%%%%%%%%%%%%%%%%%%%%%%%%%%%%%%%%%%%%%%%%%%%%%
\section{THEORETICAL PERFORMANCE ANALYSIS}
\label{perform}
The ergodic capacity and outage probability are fundamental to the development of reliable communication. In this section, we investigate these metrics for the introduced channel model (in Section \ref{system_and_channel}-\ref{channel_mod}), which allows a comprehensive analysis for both Gaussian and non-Gaussian processes under Nakagami-m fading. However, there is no exact capacity formulation for $\text{AS}\alpha\text{SN}$ channels when the noise stability is unknown or uncertain. Therefore, we use the lower bound of the capacity in \cite{Capacity_bound}, which also applies to the Shannon capacity under the Gaussian assumption $\alpha =2$.

The capacity of the $\text{AS}\alpha\text{SN}$ channel with $\alpha \in (1,2]$ is defined in \cite{Capacity_bound} by solving the optimization problem below.
\begin{equation}
\label{capdef}
\begin{split}
\max_{p(x) \in \mathcal {P}} \ \ \   & I(X;Y) \\
\text{subject to}       \ \        & \mathbb {E}[|X|] \leq P_c,
\end{split}
\end{equation}
where $\mathcal {P}$ denote the collection of Borel probability and $p(x)$ is all possible distributions for random input variables $X$. $I(X;Y)$ is the mutual information of the channel given by $Y=X+N$, where $N$ and $Y$ represent random noise and output variables, respectively. The input variables of the channel are subject to the constraint $P_c > 0$. The authors of \cite{Capacity_bound} solve the optimization problem above and prove that (\ref{capdef}) leads to the tractable lower bound of the capacity as follows
\begin{equation}
\label{c_bound}
C \geq \frac {1}{\alpha }\log _{2}\left ({1 + \left ({\frac {P_c}{\mathbb {E}[|N|]}}\right )^{\alpha }}\right ).
\end{equation} 
The study also shows analogy between the ${\frac {P_c}{\mathbb {E}[|N|]}}$ and SNR under Gaussian noise. We used the above boundary as the basic for our theoretical analysis.

As seen in Property 1, $\text{S}\alpha\text{S}$ variables for $\alpha < 2$ have no finite second-order moments, so the instantaneous received SNR under $\text{AS}\alpha\text{SN}$ is expressed by using the ${\frac {P_c}{\mathbb {E}[|N|]}}$ ratio as follows
\begin{equation}
\label{instant_snr}
\gamma = \left ( \frac{P_c \left | h \right |\pi } {2 \lambda_n \Gamma \left ({1 - \frac {1}{\alpha }}\right )}\right )^\alpha.
\end{equation}
\begin{lemma}
\label{LEMMA1}
Let $\bar{\gamma}$ be the average SNR and let $\xi = \mathbb {E}\left[ \left | h \right |^\alpha\right]$ be the expected value of $\left | h \right |^\alpha$. Then, the pdf of the instantaneous received SNR for Nakagami-m fading channel under $\text{AS}\alpha\text{SN}$, $f(\gamma)$ is derived as
\begin{equation}
\label{Naksnr_pdf}
f_{\gamma}(\gamma) = \frac{2m^m \gamma^{-1}}{ \Omega \alpha \Gamma(m)} \exp\left[-\frac{m}{\Omega} \left( \frac{\gamma\xi}{\bar{\gamma}} \right)^{\frac{2}{\alpha}}\right] \left( \frac{\gamma \xi }{\bar{\gamma}} \right)^{\frac{2m}{\alpha}}, 
\end{equation}
where $m$ denotes the Nakagami-m fading parameter and $\Omega$ is the mean-square of the fading amplitude.
\end{lemma}
\renewcommand\qedsymbol{$\blacksquare$}
\begin{proof}
The proof is presented in Appendix B.
\end{proof}
%%%%%%%%%%%%%%%%%%%%%%%%%%%%%%%%%%%%%%%
%%%%%%%%%%%%%%%%%%%%%%%%%%%%%%%%%%%%%%%
%%__%%__%%__%%__%%
%%__%%__%%__%%__%%
%%__%%__%%__%%__%%
\subsection{Lower Bounds of Ergodic Capacity}
Wireless communication channels often encounter different propagation conditions, which makes the development of communication systems to adapt the transmission rate to the instantaneous channel capacity a challenge. Therefore, communication systems are designed to operate at ergodic capacity to ensure a sustainable information rate, in other words reliable communication.

The ergodic capacity is statistical average of the mutual information depending on the fading and receiver side information information \cite{goldsmith_cap}. Then, the lower bound of the ergodic capacity for the $\text{AS}\alpha\text{SN}$ channel is written by substituting (\ref{c_bound}) as follows
\begin{equation}
\begin{split}
\label{gold_cap}
\overline{C} \ &\geq \ \mathbb{E}\left [ \frac{1}{\alpha} \log_2(1+\gamma) \right ] \\
&\geq \ \frac{1}{\alpha}  \int  \log_2(1+\gamma) f(\gamma) d\gamma.
\end{split}
\end{equation}
%%%%%%%%%%%%%%%%%%%%%%%%%%%%%%%%%%%%%%%%%%%%%%%%%%%%%%%%%%%%%%%%%%%%
%%%%%%%%%%%%%%%%%%%%%%___THEOREM_2__%%%%%%%%%%%%%%%%%%%%%%%%%%%%%%%%
%%%%%%%%%%%%%%%%%%%%%%%%%%%%%%%%%%%%%%%%%%%%%%%%%%%%%%%%%%%%%%%%%%%%
\begin{theorem}
\label{Theorem2}
The ergodic capacity for a Nakagami-m fading channel under $\text{AS}\alpha\text{SN}$ is lower bounded by using Lemma \ref{LEMMA1} as follows
\begin{equation}
\label{ergodic_bound}
\begin{split}
 \overline{C} 
&\geq \frac{ \Omega^{m-1} l}{ 2 \mathrm{ln}(2)  \Gamma(m)} \sqrt{\frac{k^{2m-3}}{(2\pi)^{2l+k-3}}}   \\
&\times G_{2l,k+2l}^{k+2l,l} \left[ 
\frac{\left( {\xi } / \bar{\gamma}  \right)^{l}}{\left( k \Omega m^\text{-1}  \right)^k} 
\left| \begin{matrix} I(l,0),I(l,1) \\  I(k,m),I(l,0),I(l,0)  \end{matrix}  \right.\right] . 
\end{split}
\end{equation}
Here, $I(\rho  , \iota )  \triangleq \iota/\rho , (\iota+1)/\rho , \ldots, (\iota+\rho -1)/\rho $ with $\iota $ an arbitrary real value, and $\frac{l}{k} = \frac{2}{\alpha}$ while $\rho , k, l \in \mathbb{Z}^+$.
\end{theorem}
%%%%%%_______Proof_Appendix C_______%%%%%%%%
%%%%%__%%__%%__%%__%%%%__%%__%%__%%__%%%%%%%
\renewcommand\qedsymbol{$\blacksquare$}
\begin{proof}
The proof is presented in Appendix C.
\end{proof}
%%%%%%%%%%%%%%%%%%%%%%%%%%%%%%%%%%
%%%%%%%%%%%%%%%%%%%%%%%%%%%%%%%%%%
%%__%%__%%__%%__%%
%%__%%__%%__%%__%%
\subsection{Upper Bounds of Outage Probability}
The ergodic capacity provides important insights for the development of reliable communication, but does not guarantee the continuity of the desired transmission rates. There may be poorer channel conditions or unexpected anomalies, which is quite possible in cislunar space, and so communication may occasionally be interrupted. Furthermore, this can cause a critical problem for core systems that use real-time control applications during cislunar missions.

The outage probability indicates the likelihood of communication failures and is analyzed to reduce the risk for robust and reliable communication. When the desired data rate is achieved with an instantaneous received SNR of greater than or equal to $\gamma_{th}$, the outage probability is upper bounded as follows
%%%%%%%%%%%%%%%%%%%%%%%%%%%%%%%%%%%%%%%%%%%%%%%%%%%%%%%%%%%%%%%%%%%%
\begin{equation}
\label{outage_general}
P_\textup{{out}}(\gamma_{th}) \leq \int_{0}^{\gamma_{th}}  p_{\gamma}(\gamma) d\gamma.
\end{equation}
%%%%%%%%%%%%%%%%%%%%%%%%%%%%%%%%%%%%%%%%%%%%%%%%%%%%%%%%%%%%%%%%%%%%
%%%%%%%%%%%%%%%%%%%%%%___THEOREM_3__%%%%%%%%%%%%%%%%%%%%%%%%%%%%%%%%
%%%%%%%%%%%%%%%%%%%%%%%%%%%%%%%%%%%%%%%%%%%%%%%%%%%%%%%%%%%%%%%%%%%%
\begin{theorem}
\label{Theorem3}
The outage probability for Nakagami-m fading channel under $\text{AS}\alpha\text{SN}$ is upper bounded by using Lemma \ref{LEMMA1} as
%%%%%%%%%%%%%%%%%%%%%%%%%%%%%%%%%%%%%%%%%%%%%%%%%%%%%%%%%%%%%%%%%%%%
\begin{equation}
\label{Pout_nakagami_definite}
P_{\textup{out}}  (\gamma) \leq     \frac{\Omega^{m-1}}{ \Gamma(m)}
\left[  \Gamma(m) -  \Gamma \left( m,\frac{  \left ( \gamma\xi /\bar{\gamma} \right )^\frac{2}{\alpha} }{\Omega m^{\text{-1}}} \right) \right].
\end{equation}
%%%%%%%%%%%%%%%%%%%%%%%%%%%%%%%%%%%%%%%%%%%%%%%%%%%%%%%%%%%%%%%%%%%%
\end{theorem}
%%%%%%%%%%%%%%%%%%%%%%%%%%%%%%%%%%%%%%%%%%%%%%%%%%%%%%%%%%%%%%%%%%%%
%%%%%%%%%%%%%%%%%%%%%%%%%%%%%%%%%%%%%%%%%%%%%%%%%%%%%%%%%%%%%%%%%%%%
\renewcommand\qedsymbol{$\blacksquare$}
%%%%%%%%%%%%%%%%%%%%%%%%%%%%%%%%%%%%%%%%%%%%%%%%%%%%%%%%%%%%%%%%%%%%
%%%%%%%%%%%%%%%%%%%%%%%%%%%%%%%%%%%%%%%%%%%%%%%%%%%%%%%%%%%%%%%%%%%%
\vspace{0.2cm}
\begin{proof}
The proof is presented in Appendix D.
\end{proof}
%%%%%%%%%%%%%%%%%%%%%%%%%%%%%%%%%%
%%%%%%%%%%%%%%%%%%%%%%%%%%%%%%%%%%
%%%%%%%%%%%%%%%%%%%%%%%%%%%%%%%%%%%%%%%%%%%%%%%%%%%%%%%%%%%%%%%%%%%%%%%%
\begin{algorithm}[!hb]
\caption{Approximate Computation of Ergodic Capacity}\label{alg}
\begin{algorithmic}
    \For{ $i \in \{1, \dots, N_h\}$ } \vspace{0.1cm}
    \State (1) Initialize $r^{(0)}(x) = \frac {1}{M_X}$ \vspace{0.1cm}
    \State (2) Initialize $\overline{C}=0,~C_{i,0} = 0,~C_{i,{-}1} = -2\epsilon$ \vspace{0.1cm}
        \While{$C_{i,n} - C_{i,n-1} > \epsilon$} \vspace{0.1cm}
        \State (1) $C_{i,n-1} = C_{i,n}$ \vspace{0.2cm}
        \State (2) $Q^{(n)}(x|y) = \frac {r^{(n-1)}(x)P(y|x)}{\sum _{x=1}^{M_X} r^{(n-1)}(x)P(y|x)}$ \vspace{0.2cm}
        \State (3) $C_{i,n} = \sum _{x = 1}^{M_X} \sum _{y=1}^{M_N} r^{(n-1)}(x)P(y|x)$ \vspace{0.2cm} 
        \State $ \qquad \qquad \times \log_{2} \left ({\frac {Q^{(n)}(x|y)}{r^{(n-1)}(x)}}\right )$ \vspace{0.2cm} 
        \State (4) Solve for $\nu$ such that
        \begin{equation} \sum_{x = 1}^{M_X} \left ({1 - \frac {|x|}{P_c {h_i}}}\right )e^{\nu |x|}\prod _{y = 1}^{M_N} Q^{(n)}(x|y)^{P(y|x)} = 0 \end{equation} \vspace{0.2cm}
        \begin{equation} \hspace{-0.05cm} r^{(n)}(x_i) = \frac {e^{\nu |x_i|}\prod_{y=1}^{M_N} Q^{(n)}(x|y)^{P(y|x)}}{\sum_{x^{'} = 1}^{M_X} e^{\nu |x^{'}|}\prod_{y=1}^{M_N} Q^{(n)}(x^{'}|y)^{P(y|x^{'})}} \end{equation} 
        \EndWhile
        \State \Return $C_{i,n} $\vspace{0.15cm} 
    \State (5) \label{step5} $\overline{C}= \overline{C} + C_{i,n} f_{{\gamma}}(\gamma_i) \Delta_i $\vspace{0.15cm} 
    \EndFor \vspace{0.1cm} 
    \State \Return $\overline{C}$
\end{algorithmic}
\end{algorithm}
%%__%%__%%__%%__%%
%%__%%__%%__%%__%%
%%__%%__%%__%%__%%
%%__%%__%%__%%__%%
%%%%%%%%%%%%%%%%%%%%%%%%%%%%%%%%%%%%%%%%%%%%%%%%%%%%%%%%%%%%%%%%%%%%%%%%
%%%%%%%%%%%%%%%%%%%%%%%%%%%%%%%%%%%%%%%%%%%%%%%%%%%%%%%%%%%%%%%%%%%%%%%%
%%%%%%%%%%%%%%%%%%%%%%%%%%%%%%%%%%%%%%%%%%%%%%%%%%%%%%%%%%%%%%%%%%%%%%%%
%%%%%%%%%%%%%%%%%%%%%%%%%%%%%%%%%%%%%%%%%%%%%%%%%%%%%%%%%%%%%%%%%%%%%%%%
%%%%%%%%%%%%%%%%%%%%%%%%%%%%%%%%%%%%%%%%%%%%%%%%%%%%%%%%%%%%%%%%%%%%%%%%
%%%%%%%%%%%%%%%%%%%%%%%%%%%%%%%%%%%%%%%%%%%%%%%%%%%%%%%%%%%%%%%%%%%%%%%%
%%%%%%%%%%%%%%%%%%%%%__Numerical Analysis__%%%%%%%%%%%%%%%%%%%%%%%%%%%%%
%%%%%%%%%%%%%%%%%%%%%%%%%%%%%%%%%%%%%%%%%%%%%%%%%%%%%%%%%%%%%%%%%%%%%%%%
%%%%%%%%%%%%%%%%%%%%%%%%%%%%%%%%%%%%%%%%%%%%%%%%%%%%%%%%%%%%%%%%%%%%%%%%
%%%%%%%%%%%%%%%%%%%%%%%%%%%%%%%%%%%%%%%%%%%%%%%%%%%%%%%%%%%%%%%%%%%%%%%%
%%%%%%%%%%%%%%%%%%%%%%%%%%%%%%%%%%%%%%%%%%%%%%%%%%%%%%%%%%%%%%%%%%%%%%%%
%%%%%%%%%%%%%%%%%%%%%%%%%%%%%%%%%%%%%%%%%%%%%%%%%%%%%%%%%%%%%%%%%%%%%%%%
\subsection{Numerical Approximation for Ergodic Capacity}\label{blahut}
The theoretical bounds for the ergodic capacity are presented in closed form under $\text{AS}\alpha\text{SN}$ and Nakagami-m fading. Since we perform the derivations with closed-form expressions, the bound in Theorem \ref{Theorem3} preserves the tightness of the basic bound in equation (\ref{c_bound}). However, in this section, we propose to show the tightness of the bound to give a clear insight into our results in the following section. We compare the lower bound of the ergodic capacity with the numerically approximated capacity values using the Blahut-Arimoto algorithm for the $\text{AS}\alpha\text{SN}$ in \cite{Capacity_bound}.

The Blahut-Arimoto algorithm is a well-known capacity approximation algorithm for discrete memoryless channels where the inputs must be discrete and with a finite alphabet. To the best of our knowledge, this is the first in the literature where the Blahut-Arimoto algorithm has been adapted for the presence of fading, as detailed in Algorithm \ref{alg}. We defined the channel as follows
\begin{equation}
Y = h_i X + N, \ \ i = 1, 2, \cdots, N_h,
\end{equation}
where $h_i$ is a fading coefficient, $X$ and $N$ are random variables with finite alphabets $S_{X}$ and $S_{N}$ with lengths $M_X$ and $M_N$, respectively. In Algorithm \ref{alg}, we compute the capacity values for each $h_i$ and then obtained the approximation to the ergodic capacity using Riemann summation \cite{calculus} in step (5). The rest of the algorithm is implemented in the same way as in \cite{Capacity_bound}. 

\begin{figure}[!hb]
\centering
\includegraphics[width=\linewidth]{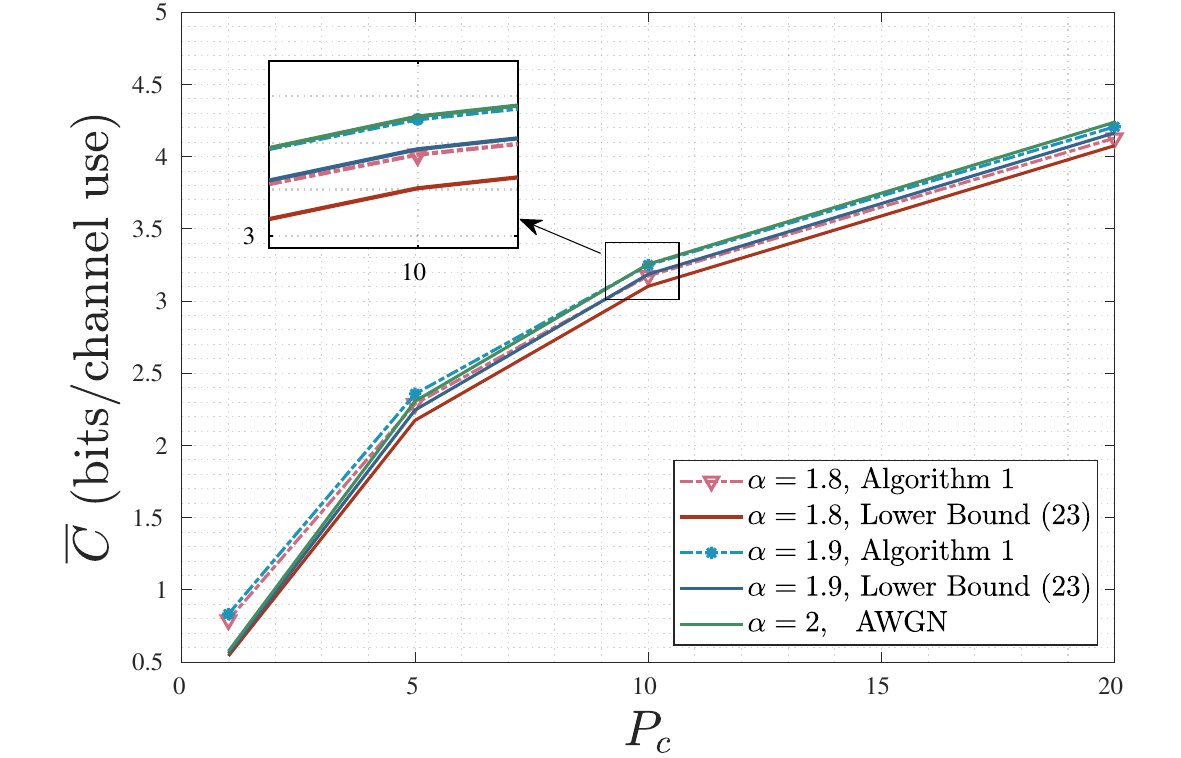}
\caption{Comparisons of the closed-form lower bound for ergodic capacity in equation (\ref{ergodic_bound}) and numeric capacity approximation algorithm with $m=1$ and $\alpha \in \left \{1.8, 1.9, 2\right \}$.}
\label{ba_m1}
\end{figure}
\begin{figure}[!ht]
\centering
\includegraphics[width=\linewidth]{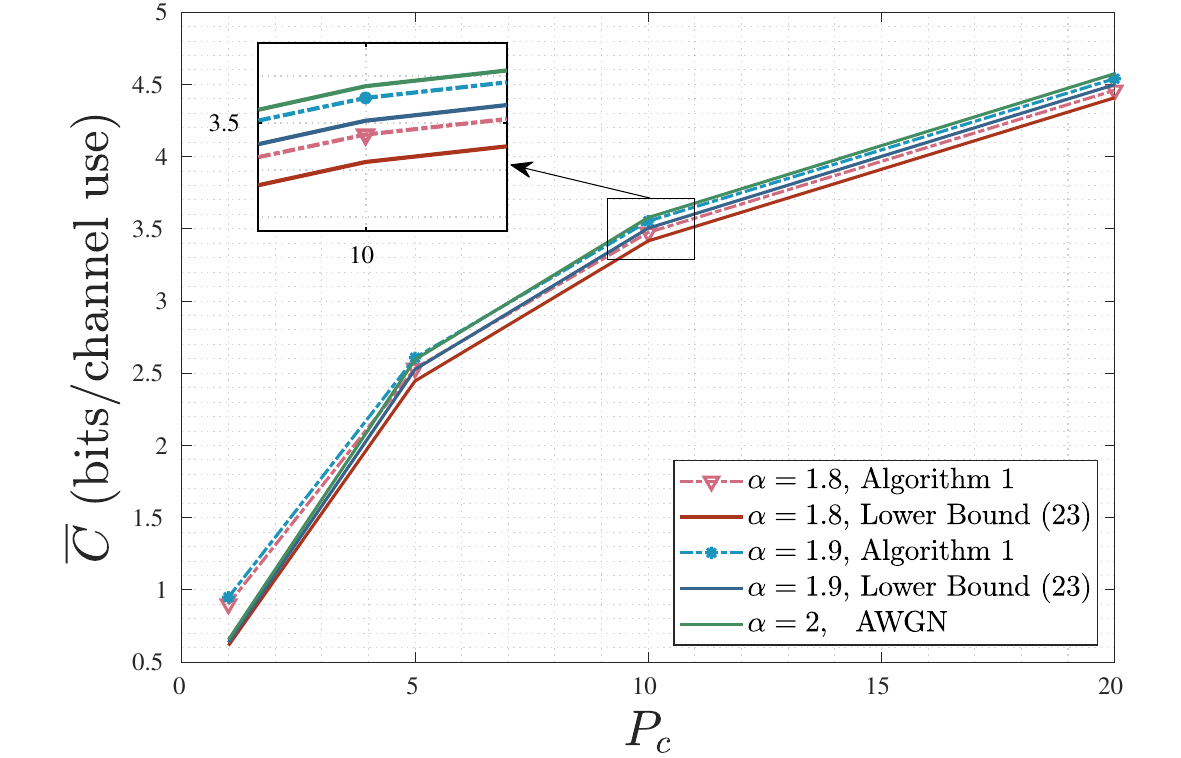}
\caption{Comparisons of the closed-form lower bound for ergodic capacity in equation (\ref{ergodic_bound}) and numeric capacity approximation algorithm with $m=5$ and $\alpha \in \left \{1.8, 1.9, 2\right \}$.}
\label{ba_m5}
\end{figure}
\begin{figure}[!htb]
\centering
\includegraphics[width=\linewidth]{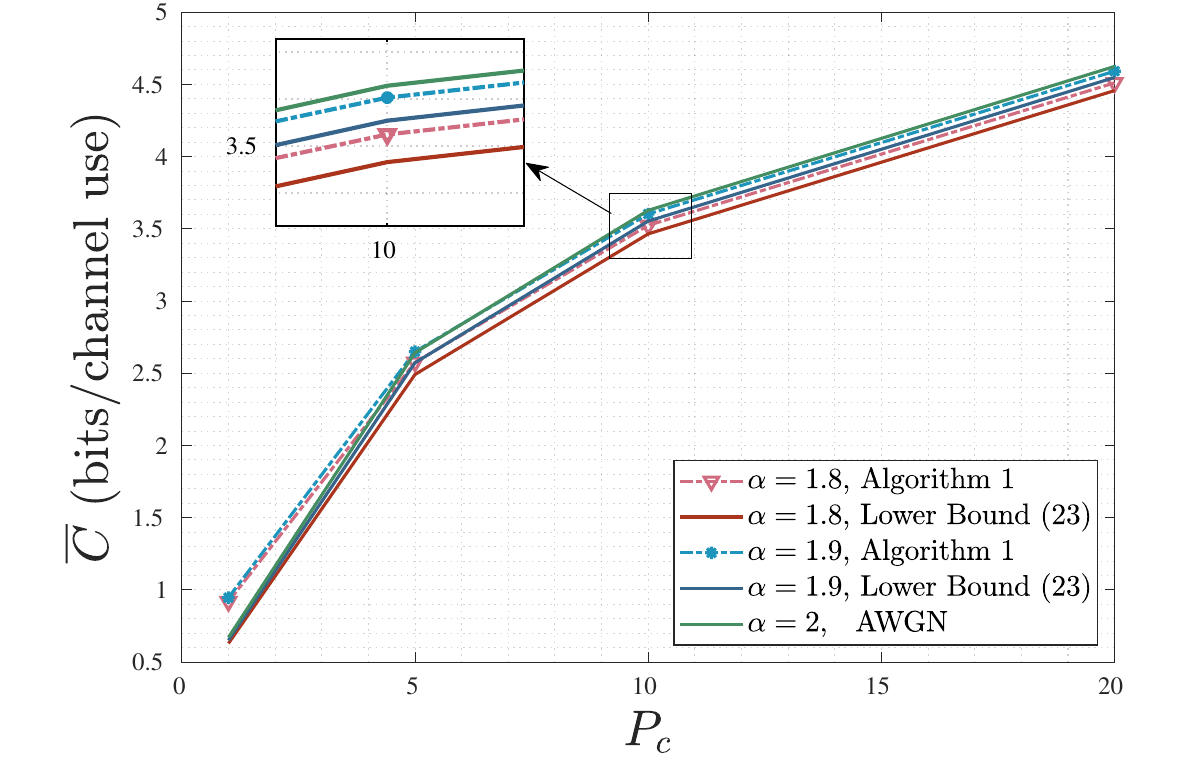}
\caption{Comparisons of the closed-form lower bound for ergodic capacity in equation (\ref{ergodic_bound}) and numeric capacity approximation algorithm with $m=15$ and $\alpha \in \left \{1.8, 1.9, 2\right \}$.}
\label{ba_m15}
\end{figure}

%%%%%%%%%%%%%%%%%%%%%%%%%%%%%%%%%%%%%%%%%%%%%%%
Figures \ref{ba_m1}, \ref{ba_m5} and \ref{ba_m15} demonstrate the tightness of the closed-form lower bound in equation (\ref{ergodic_bound}) for three fading and stability conditions individually while channel inputs are subject to the constraint $P_c \in \{1, 5, 10, 20\}$. The results also include the ergodic capacity for AWGN channel that can also be considered $\text{AS}\alpha\text{SN}$ channel with $\alpha=2$. All results are obtained by setting the shift parameter of noise variables $\lambda = 1/\sqrt{2}$ that is identical with unit variance for noise variables in AWGN channel as proven in Appendix A. 

We compare the lower bound and numerical approximation of ergodic capacity in Fig. \ref{ba_m1} for Nakagami-m fading parameter $m=1$ that corresponds to Rayleigh fading or Rician fading with $K = 0$. Fig. \ref{ba_m1} indicates the impact of $\alpha$ values on the ergodic capacity and the compatibility of the lower bound in equation (\ref{ergodic_bound}). For example, under the constraint of $P_c=5$, the approximate results of the ergodic capacity are 2.2883 bits per channel use (bpcu) and 2.3573 bpcu for $\alpha=1.8$ and $\alpha=1.9$, respectively while the ergodic capacity is 2.3066 bpcu for $\alpha=2$.
%%%%%%%%%%%%%%%%%%%%%%%%%%%%%%%%%%%%%%%%%%%%%%%
%%%%%%%%%%%%%%%%%%%%%%%%%%%%%%%%%%%%%%%%%%%%%%%

Fig. \ref{ba_m5} shows the results for better fading condition with $m=5$, which can also be considered for Rician fading with $K \approx 8.5$. In Fig. \ref{ba_m5}, we observe that the lower bound are packed with the approximate average capacity tightly and how the ergodic capacity increases with improved propagation and noise conditions. 
%%%%%%%%%%%%%%%%%%%%%%%%%%%%%%%%%%%%%%%%%%%%%%%
%%%%%%%%%%%%%%%%%%%%%%%%%%%%%%%%%%%%%%%%%%%%%%%

The same comparison is presented in Fig. \ref{ba_m15} to show the results for mild fading with Nakagami-m fading parameter $m=15$. The closeness between the closed-form lower bound and the numerical approximation for capacity does not alter depending on m parameter. In the zoomed window, for $\alpha=1.9$ and $P_c=10$, the numerical approximation of the ergodic capacity by Algorithm \ref{alg} is 3.6028 bpcu, while the theoretical bound by (\ref{ergodic_bound}) is 3.5536 bpcu. For $\alpha=1.8$ and $P_c=10$, these results are 3.5242 and 3.4652 bpcu, respectively. The differences between the numerical approximation and the bound are 0.0492 and 0.059 for $\alpha=1.9$ and $\alpha=1.8$, respectively. We observe that the tightness of the bound decreases slightly with decreasing alpha values, just as with the basic bound in \cite{Capacity_bound}. Overall, the results show that the lower bound for ergodic capacity is tighter in the case $\alpha=1.9$ compare to $\alpha=1.8$ exactly like to the behavior\footnote[5]{\cite{Capacity_bound} shows that the tightness of (\ref{c_bound}) decreases on average by 1 bit for $\alpha=1.1$ and that it is consistent with our results for $\alpha=1.9$. Note that we do not consider the highly extreme cases in this study that disrupt the channel stability this level.} of the basic bound in (\ref{c_bound}). This also shows that closed-form expressions in Section \ref{perform} do not affect the tightness of (\ref{c_bound}) but also the performance of Algorithm 1.
%%%%%%%%%%%%%%%%%%%%%%%%%%%%%%%%%%%%%%%%%%%%%%%%%%%%%%%%%%%%%%%%%%%%%%%%
%%%%%%%%%%%%%%%%%%%%%%%%%%%%%%%%%%%%%%%%%%%%%%%%%%%%%%%%%%%%%%%%%%%%%%%%
%%%%%%%%%%%%%%%%%%%%%%%%%%%%%%%%%%%%%%%%%%%%%%%%%%%%%%%%%%%%%%%%%%%%%%%%
%%%%%%%%%%%%%%%%%%%%%%%%%%%%%%%%%%%%%%%%%%%%%%%%%%%%%%%%%%%%%%%%%%%%%%%%
%%%%%%%%%%%%%%%%%%%%%%%%%%%%%%%%%%%%%%%%%%%%%%%%%%%%%%%%%%%%%%%%%%%%%%%%
%%%%%%%%%%%%%%%%%%%%%%%%%%%%%%%%%%%%%%%%%%%%%%%%%%%%%%%%%%%%%%%%%%%%%%%%
%%%%%%%%%%%%%%%%%%%%________Results________%%%%%%%%%%%%%%%%%%%%%%%%%%%%%
%%%%%%%%%%%%%%%%%%%%%%%%%%%%%%%%%%%%%%%%%%%%%%%%%%%%%%%%%%%%%%%%%%%%%%%%
%%%%%%%%%%%%%%%%%%%%%%%%%%%%%%%%%%%%%%%%%%%%%%%%%%%%%%%%%%%%%%%%%%%%%%%%
%%%%%%%%%%%%%%%%%%%%%%%%%%%%%%%%%%%%%%%%%%%%%%%%%%%%%%%%%%%%%%%%%%%%%%%%
%%%%%%%%%%%%%%%%%%%%%%%%%%%%%%%%%%%%%%%%%%%%%%%%%%%%%%%%%%%%%%%%%%%%%%%%
%%%%%%%%%%%%%%%%%%%%%%%%%%%%%%%%%%%%%%%%%%%%%%%%%%%%%%%%%%%%%%%%%%%%%%%%
%%%%%%%%%%%%%%%%%%%%%%%%%%%%%%%%%%%%%%%%%%%%%%%%%%%%%%%%%%%%%%%%%%%%%%%%
\begin{table}[!hb]
\centering
\caption{Link budget parameters for the communication links \cite{ref1_t1,ref2_t1,ref5_t1,ref_freqs}, as shown in Fig. \ref{SYSTEM}.}
\resizebox{\columnwidth}{!}{%
\begin{tblr}{
  cells = {c},
  cell{1}{1} = {c=2,r=2}{},
  cell{1}{3} = {r=2}{},
  cell{1}{4} = {c=2}{},
  cell{3}{1} = {c=2}{},
  cell{4}{1} = {c=2}{},
  cell{5}{1} = {r=5}{},
  cell{5}{4} = {c=2}{},
  cell{6}{4} = {c=2}{},
  cell{7}{4} = {c=2}{},
  cell{9}{4} = {c=2}{},
  cell{10}{1} = {r=9}{},
  cell{10}{4} = {c=2}{},
  cell{11}{4} = {c=2}{},
  cell{13}{4} = {c=2}{},
  cell{14}{4} = {c=2}{},
  cell{15}{4} = {c=2}{},
  cell{16}{4} = {c=2}{},
  cell{17}{4} = {c=2}{},
  cell{18}{4} = {c=2}{},
  cell{19}{1} = {c=2}{},
  cell{19}{4} = {c=2}{},
  vlines,
  hline{1,3-5,10,19-20} = {-}{},
  hline{2} = {4-5}{},
  hline{6-9,11-18} = {2-5}{},
}
\textbf{Parameters}                                &              & \textbf{Unit} & \textbf{Lunar Surface to Lunar Gateway} &        \\
                                                   &              &               &  \ \ \  \textbf{S Band} \ \ \       & \textbf{Ka Band} \\
$f$                                                &              & MHz           & 2245                                    & 27250  \\
Bandwidth                                          &              & MHz           & 1                                       & 10     \\
\begin{sideways}\textbf{Transmitter}\end{sideways} 
                                                   & $P_t$        & W             & 1, 10                                   &        \\
                                                   & $\eta_{A,t}$ & \%            & 43                                      &        \\
                                                   & $D_t$        & m             & 0.254                                   &        \\
                                                   & $G_t$        & dBi           & 11.85                                   & 28.27  \\
                                                   & $L_t$        & dB            & 1                                       &        \\
\begin{sideways}\textbf{Receiver}\end{sideways}    
                                                   & $\eta_{A,r}$ & \%            & 54                                      &        \\
                                                   & $D_r$        & m             & 1.5                                     &        \\
                                                   & $G_r$        & dBi           & 33.53                                   & 49.95  \\
                                                   & $L_r$        & dB            & 3                                       &        \\
                                                   & $T_{AP}$     & K             & 300                                     &        \\
                                                   & $T_R$        & K             & 50                                      &        \\
                                                   & $T_{TLP}$    & K             & 300                                     &        \\
                                                   & $\eta_{rad}$ & \%            & 95                                      &        \\
                                                   & $\eta_{TL}$  & \%            & 99                                      &        \\
$T_{CMB}$                                          &              & K             & 2.725                                   &        
\end{tblr}
}
\label{Linkbudparams}
\end{table}
%%%%%%%%%%%%%%%%%%%%%%%%%%%%%
%%%%%%%%%%%%%%%%%%%%%%%%%%%%% 
%%%%%%%%%%%%%%%%%%%%%%%%%%%%%
\section{PERFORMANCE ANALYSIS AT SYSTEM LEVEL}\label{results}
%%%%%%%%%%%%%%%%%%%%%%%%%%%%%
%%%%%%%%%%%%%%%%%%%%%%%%%%%%%
%%%%%%%%%%%%%%%%%%%%%%%%%%%%%
%%%%%%%%%%%%%%%%%%%%%%%%%%%%%%%%%%%%%%%%%%%%%%
\subsection{CSN Architecture: Lunar Gateway}
We propose the cislunar communication through the communication link of Lunar Gateway, as shown in Fig. \ref{SYSTEM}. It is orbiting in NRHO, unlike others, and is a far preferable relay orbiter as its presence is projected at least 15 years \cite{ref1_t1}. Another key factor in the popularity of Lunar Gateway is its design, which aims to provide broader dynamic connectivity with maximum interoperability \cite{ref2_t1}. It is capable of providing both low and high data rate mission services for users on the lunar surface or in orbit. Thanks to its eligibility for mission objectives, Lunar Gateway designed to support not only the most critical Artemis missions, many lunar missions including commercial payload services and inter-agency purposes, but also future deep space missions \cite{Intro4,ref1_t1,ref2_t1}. Therefore, it is also seen as an important attempt to determine the future standards of CSNs.

For the communication scenario considered in this study, as shown in Fig. \ref{SYSTEM}, Lunar Gateway is the receiver and the transmitter is a rover on the lunar surface, which is also the communication terminal of the mission. The communication terminal transmits the communication signals over S band (2200 - 2290 MHz) or Ka band (27 - 27.5 GHz) \cite{ref2_t1}. The targeted maximum data rates for the link from lunar rover to the Lunar Gateway are 4 Msps and 100 Mbps in S and Ka bands \cite{ref_freqs}, respectively. In accordance with the mission objectives and the use case of bands, we determine the bandwidths for each band, as shown in Table \ref{Linkbudparams}.
%%%%%%%%%%%%%%%%%%%%%%%%%%%%%___Ka*BAND___%%%%%%%%%%%%%%%%%%%%%%%%%%%%%
%%__%%__%%__%%__%%
%%__%%__%%__%%__%%
%%__%%__%%__%%__%%
%%__%%__%%__%%__%%
\begin{figure*}[!hb]
\centering
\begin{subfigure}{\columnwidth}
\centering
\includegraphics[width=\linewidth]{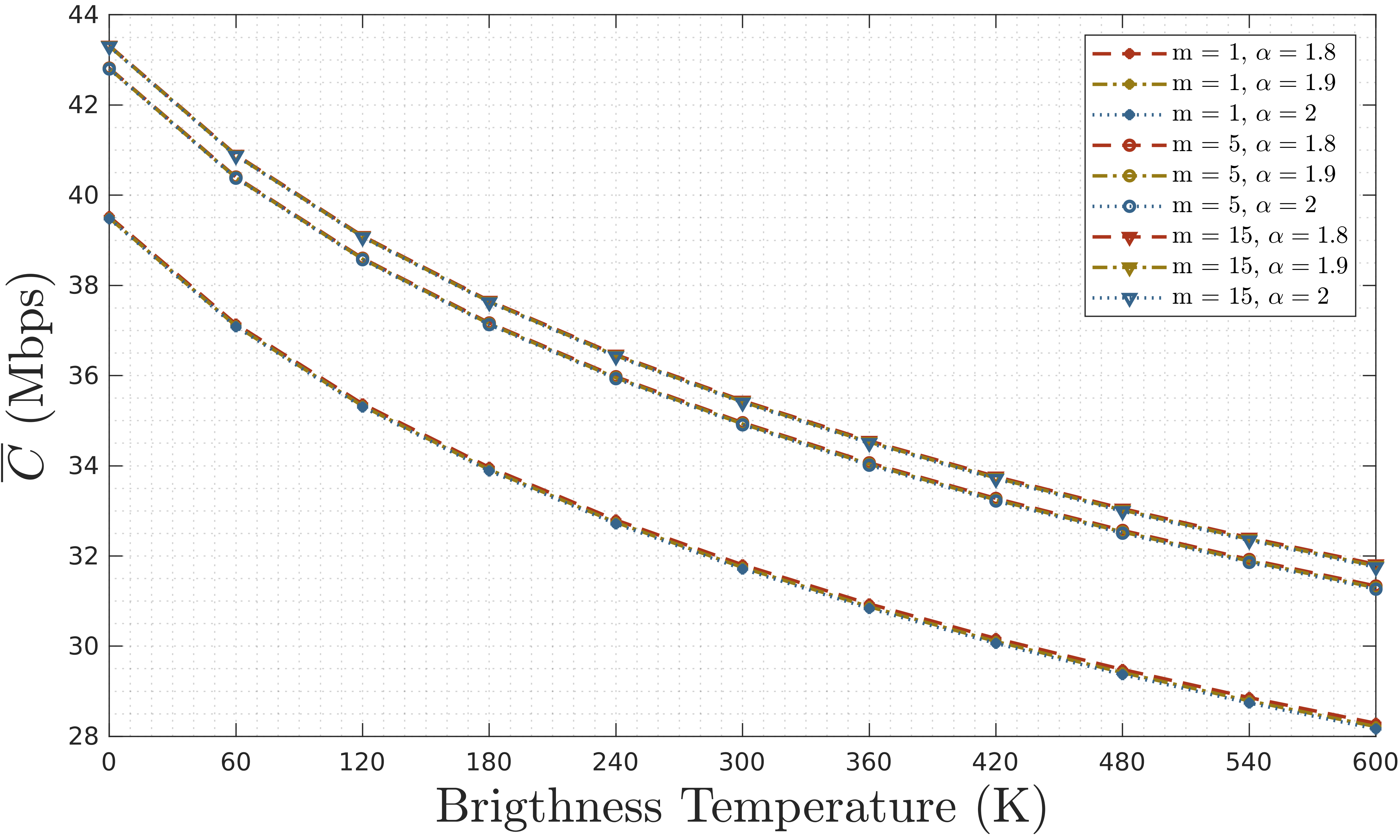}
\caption{$d = 10\times10^6$ m, $P_t = 1$ W.}
\label{KA_C_dmin_Pt1}
\end{subfigure}\hspace*{\fill}
\begin{subfigure}{\columnwidth}
\centering
\includegraphics[width=\linewidth]{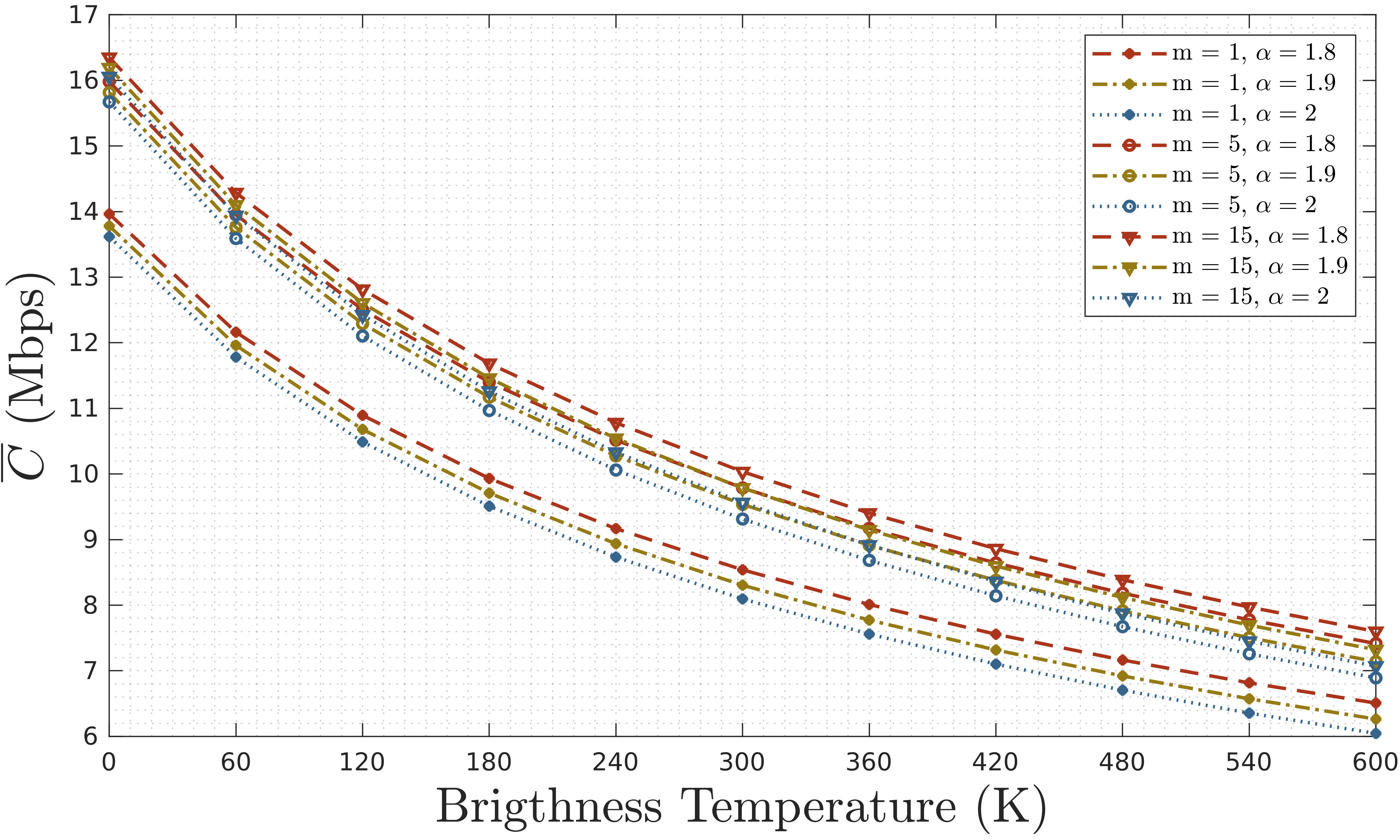}
\caption{$d = 70\times10^6$ m, $P_t = 1$ W.}
\label{KA_C_dmax_Pt1}
\end{subfigure}\hspace*{\fill}
\\
\begin{subfigure}{\columnwidth}
\centering
\includegraphics[width=\linewidth]{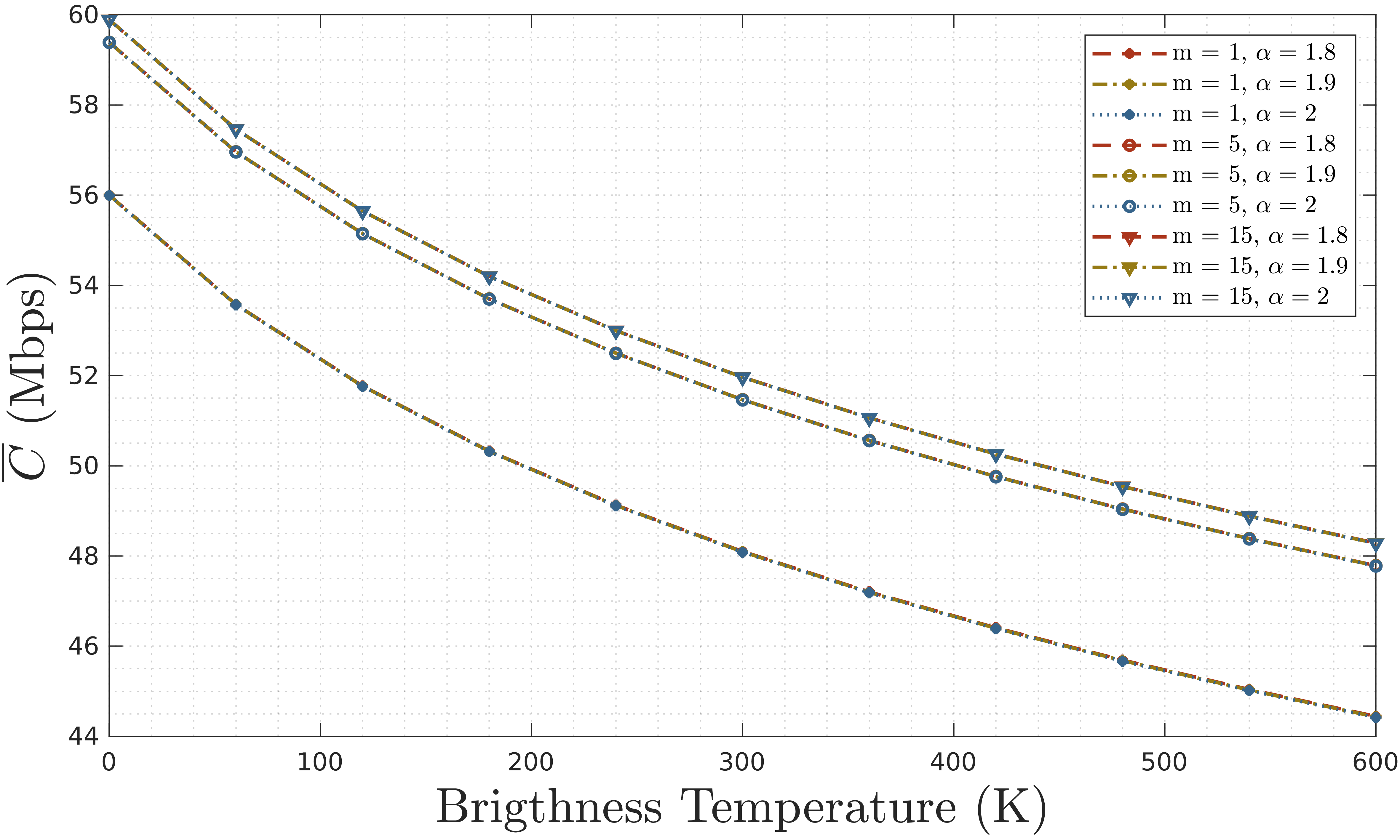}
\caption{$d = 10\times10^6$ m, $P_t = 10$ W.}
\label{KA_C_dmin_Pt10}
\end{subfigure}\hspace*{\fill}
\begin{subfigure}{\columnwidth}
\centering
\includegraphics[width=\linewidth]{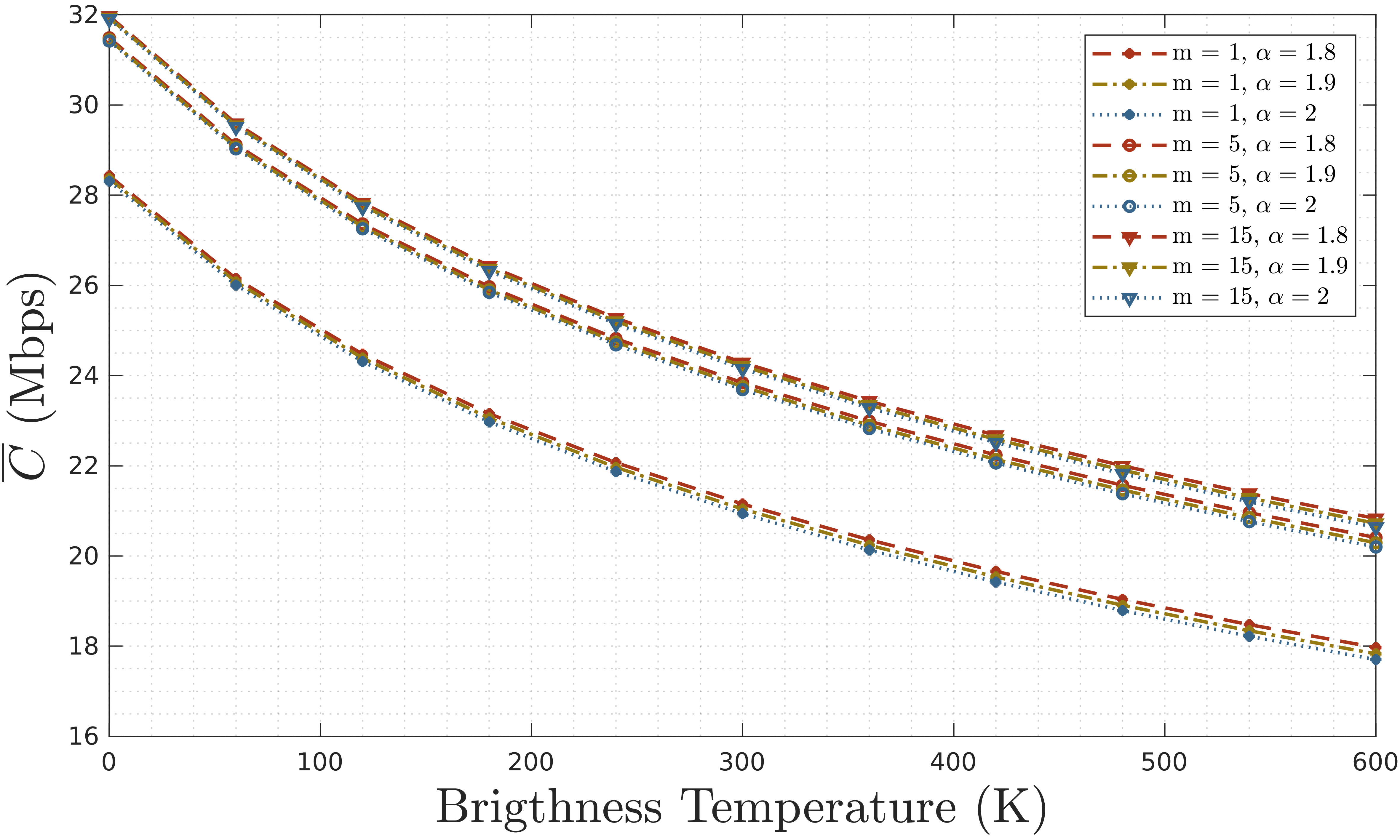}
\caption{$d = 70\times10^6$ m, $P_t = 10$ W.}
\label{KA_C_dmax_Pt10}
\end{subfigure}
\caption{The comparison of the ergodic capacity for the Ka bandwidth the lower bound in equation (\ref{ergodic_bound}) depending on the brightness temperature and under different system settings.}
\label{KA_C}
\end{figure*}
%%__%%__%%__%%__%%
%%__%%__%%__%%__%%
%%__%%__%%__%%__%%
%%__%%__%%__%%__%%
\subsection{Simulation Set-up and Configuration}\label{sim_setup}
All theoretical and system analyzes are performed through intertwined results using the parameters in Table \ref{Linkbudparams}. By diversifying the worse and better conditions for different mission objectives, we aim to provide a broader insight into the development and optimization of robust cislunar communication. Before we present the results for the ergodic capacity and the outage probability, we explain for the system set-up and configurations below.
\begin{itemize}
%%%
\item The brightness temperature of Moon changes daily as the Moon, Earth and sun orbit each other. As the Lunar Gateway's antenna is pointed at the transmitter on the lunar surface, the noise temperature of the receiver changes depending on $T_B$. Our results cover the range of $T_B$ from 0 to 600 K.
%%%
\item While the Lunar Gateway is moving in NRHO and pointing the antenna at a lunar rover, the lunar rover is also moving between sunlit and shadowed areas. As a result of these and other uncertainties, the noise temperature (so the communication signals) can show an impulsive behavior. With this in mind, we extend the results for three different conditions of noise stability with $\alpha \in \left \{1.8, 1.9, 2\right \}$.
%%%
\item During the missions, the propagation mechanisms vary due to the arrival angles of the signal or multifaceted and dynamic nature of the lunar surface. Our results include three fading conditions with $m \in \left \{1, 5, 15\right \}$.
%%%
\item The results are extended for the S and Ka band, taking into account mission targets with low and high data rates. The simulations are performed for center frequency of each band \cite{ref1_t1}.
%%%
\item Due to the orbital motion of the Lunar Gateway, the distance ($d$) between the transmitter and receiver changes considerably and thus also the path loss and the SNR. Therefore, the results are presented for minimum ($10\times10^6$ m) and maximum ($70\times10^6$ m) distances to observe edge cases.
%%%
\item By considering power constraints, we analyze the communication link for two transmit power $P_t \in \left \{1, 10\right \}$.
%%%
\end{itemize}
%%%%%%%%%%%%%%%%%%%%
%%%%%%%%%%%%%%%%%%%%
%%%%%%%%%%%%%%%%%%%%
%%%%%%%%%%%%%%%%%%%%
%%%%%%%%%%%%%%%%%%%%
%%%%%%%%%%%%%%%%%%%%
%%%%%%%%%%%%%%%%%%%%
%%%%%%%%%%%%%%%%%%%%
%%%%%%%%%%%%%%%%%%%%
%%%%%%%%%%%%%%%%%%%%
%%%%%%%%%%%%%%%%%%%%
%%%%%%%%%%%%%%%%%%%%
\subsection{Ergodic Capacity Lower Bound}\label{ergodiccapacity_results}
The results for the lower bound of the ergodic capacity are presented separately for each band in this section. As mentioned, the system settings are varied to investigate the susceptibility of cislunar communication and to gain insights into the design and development of future CSNs.
%%__%%__%%__%%__%%%%__%%__%%__%%__%%%%__%%__%%__%%__%%%%__%%__%%__%%__%%%%__%%__%%__%%__%%
\begin{figure*}[!b]
\centering
\begin{subfigure}{\columnwidth}
\centering
\includegraphics[width=\linewidth]{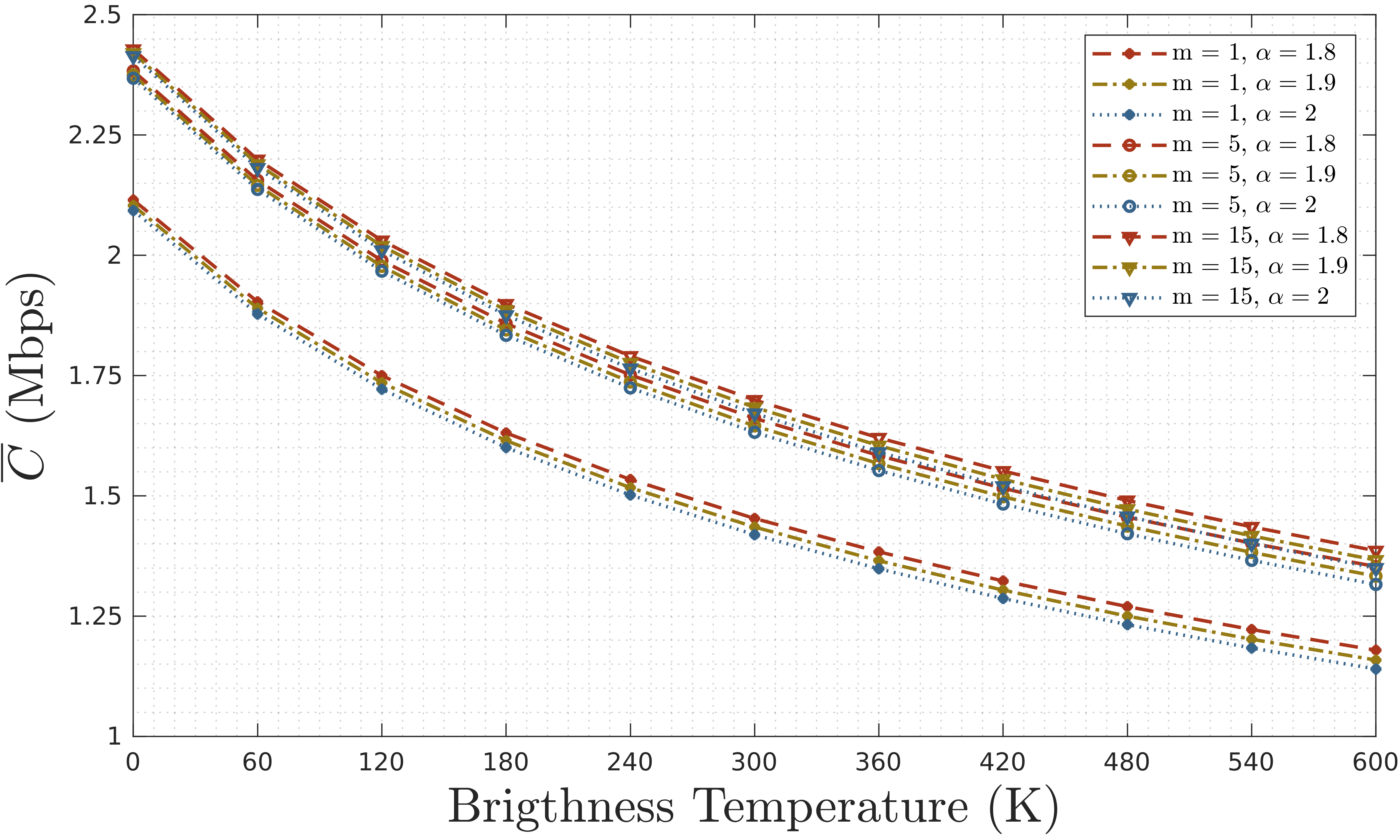}
\caption{$d = 10\times10^6$ m, $P_t = 1$ W.}
\label{S_C_dmin_Pt1}
\end{subfigure}\hspace*{\fill}
\begin{subfigure}{\columnwidth}
\centering
\includegraphics[width=\linewidth]{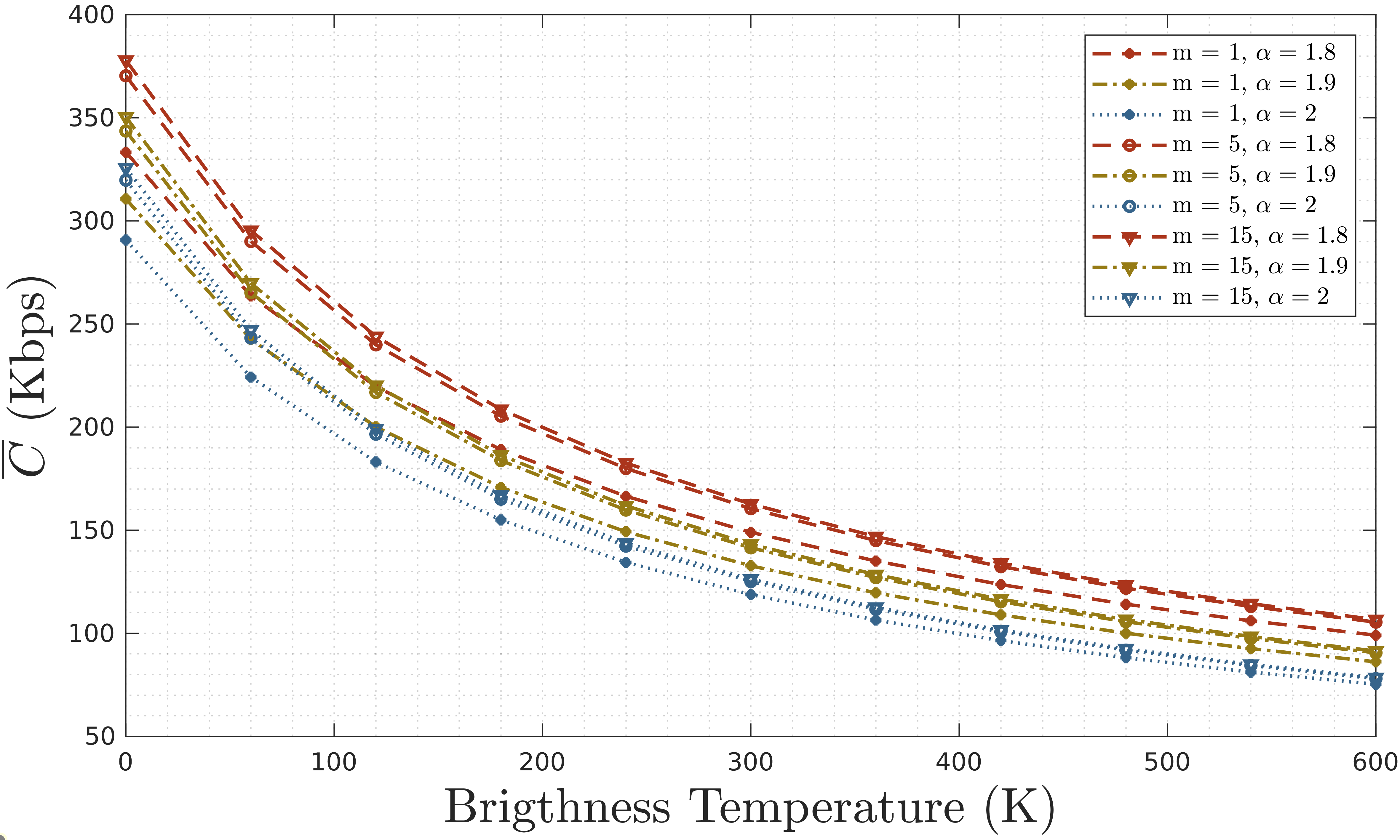}
\caption{$d = 70\times10^6$ m, $P_t = 1$ W.}
\label{S_C_dmax_Pt1}
\end{subfigure}\hspace*{\fill}
\\
\begin{subfigure}{\columnwidth}
\centering
\includegraphics[width=\linewidth]{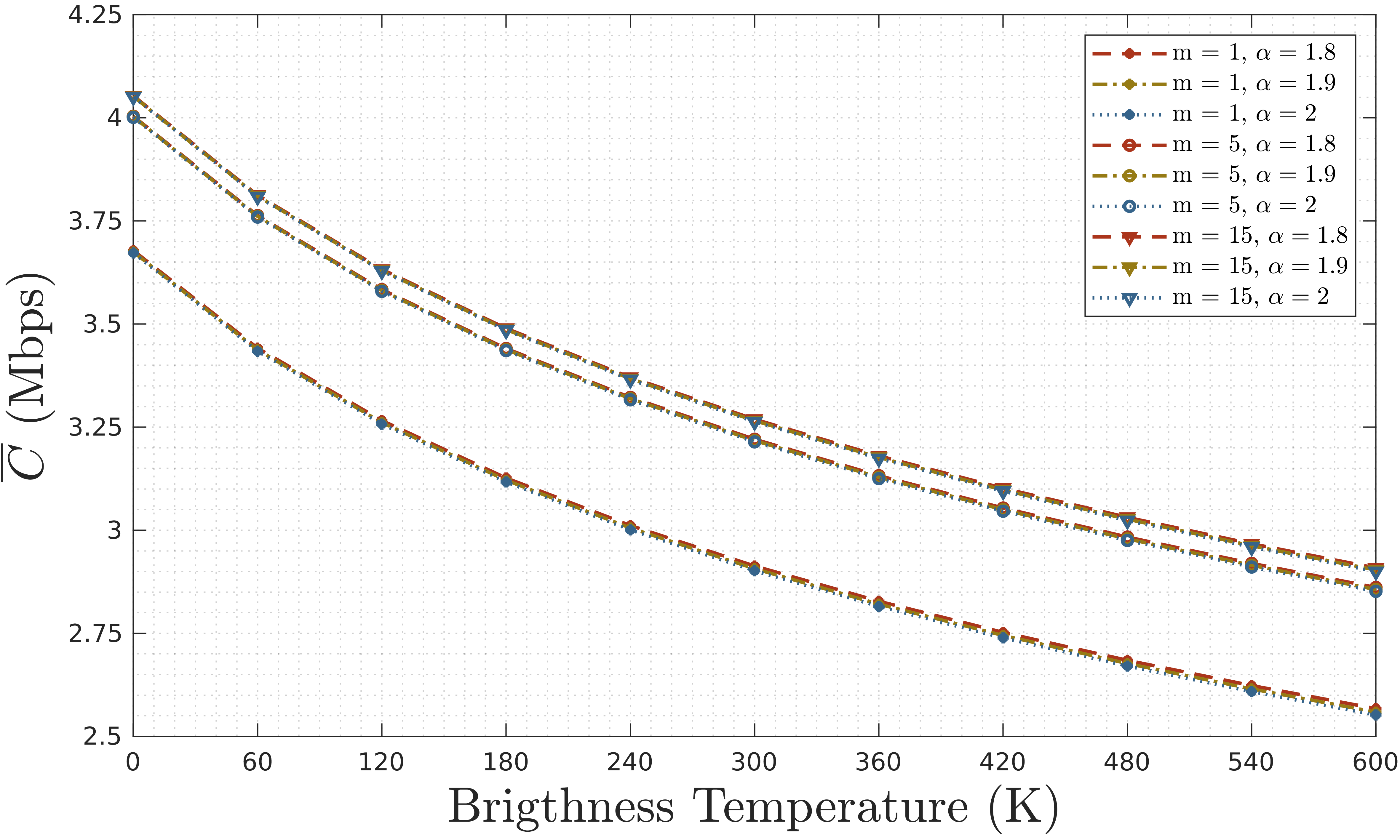}
\caption{$d = 10\times10^6$ m, $P_t = 10$ W.}
\label{S_C_dmin_Pt10}
\end{subfigure}\hspace*{\fill}
\begin{subfigure}{\columnwidth}
\centering
\includegraphics[width=\linewidth]{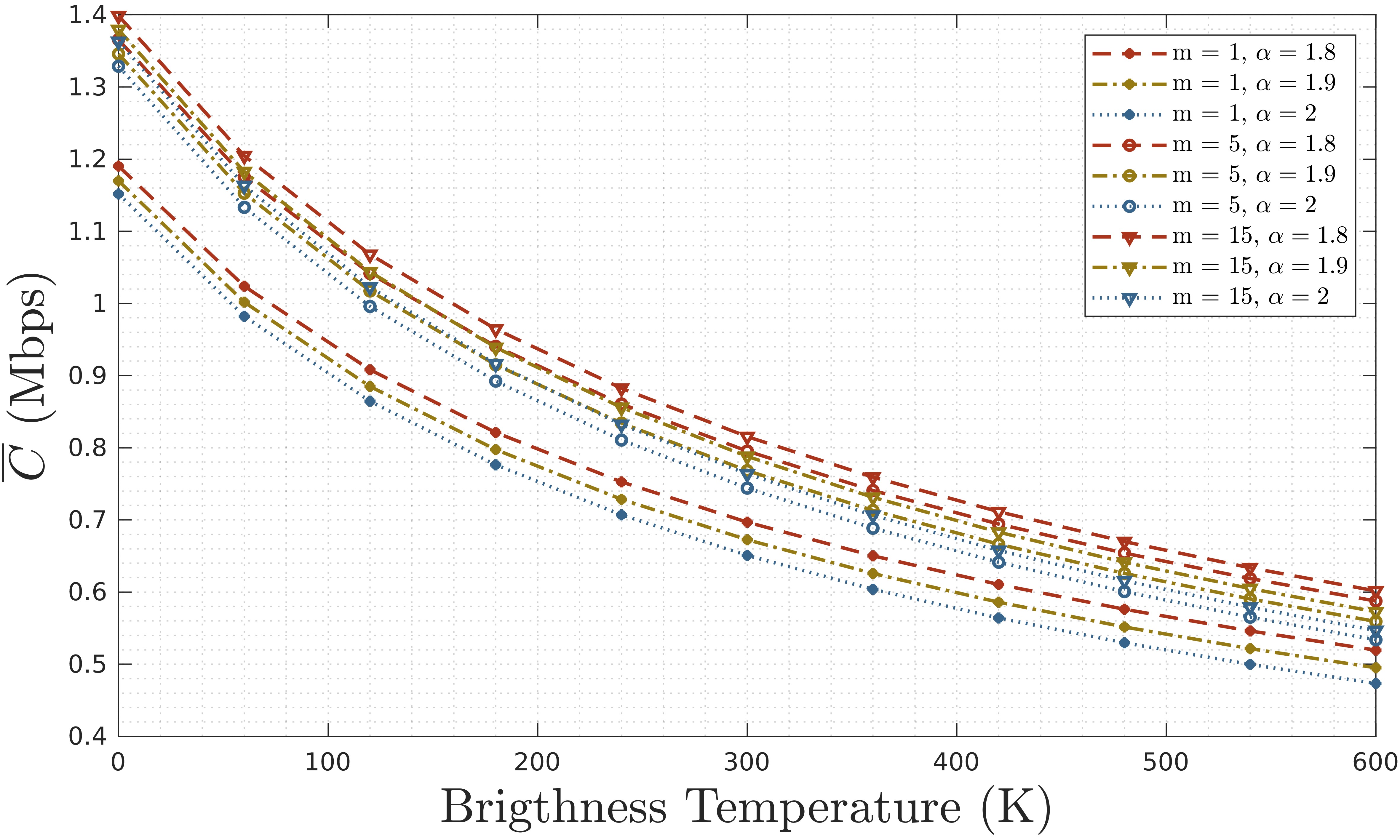}
\caption{$d = 70\times10^6$ m, $P_t = 10$ W.}
\label{S_C_dmax_Pt10}
\end{subfigure}
\caption{The comparison of the ergodic capacity for the S bandwidth the lower bound in equation (\ref{ergodic_bound}) depending on the brightness temperature and under different system settings.}
\label{S_C}
\end{figure*}
%%__%%__%%__%%__%%
%%__%%__%%__%%__%%
%%__%%__%%__%%__%%
%%__%%__%%__%%__%%

%%__%%__%%__%%__%%%%__%%__%%_______KA BAND CAPACITY_____%%%%__%%__%%__%%__%%%%__%%__%%__%%__%%
Fig. \ref{KA_C} shows the relationship between the brightness temperature and the lower bound of the ergodic capacity for Ka band under different combinations of system and channel parameters, such as the distance between the transmitter (Tx) and receiver (Rx), the transmit power, the noise stability index ($\alpha$-values) and the fading conditions. The lower bound of the ergodic capacity varies considerably and ranges from 6 Mbps to 60 Mbps.

Fig. \ref{KA_C_dmin_Pt1} presents the results for the minimum Tx-Rx distance and the lowest transmit power. The lower bound of the ergodic capacity decreases significantly from 43.5 Mbps to 28 Mbps when the brightness temperature increases from 0 K to 600 K. However, changes in noise stability ($\alpha$-values) have no significant influence on the ergodic capacity. Instead, it is the fading conditions that cause differences in the ergodic capacity, about 15.5 Mbps, when the brightness temperature increases. At the maximum Tx-Rx distance with the lowest $P_t$, as shown in Fig. \ref{KA_C_dmax_Pt1}, the ergodic capacity decreases further, from 6 Mbps to 16.5 Mbps, which is due to the increased path loss associated with the longer distance. The effect of $\alpha$-values becomes more pronounced compared to the results in Fig. \ref{KA_C_dmin_Pt1}, but worse noise conditions with lower $\alpha$-values do not lead to a decrease in average capacity. This is because the $\alpha$-values are kept constant, while the noise power changes with the brightness temperature. Consequently, the effect of increasing $T_B$ is reflected in the noise scale parameter more than the noise stability index, leading to $\lambda_{\alpha=2} >\lambda_{\alpha=1.9} > \lambda_{\alpha=1.8}$. In other words, the ergodic capacity becomes more sensitive to brightness temperature at a high noise stability index, as it leads to larger changes in $\lambda$-values. \textit{Increasing bandwidth is an effective way to reduce the impact of $\lambda$-values on ergodic capacity, but this may not be possible due to bandwidth constraints.} \textit{Therefore, an accurate estimation of the $\lambda$-value is a mandatory requirement for the design of communication systems. But the estimation of the $\alpha$-value must also be correct, as it is a prerequisite for the accurate estimation of the $\lambda$-value. In this way, we ensure that the noise scale parameter and thus the average capacity performance are maintained at a lower sensitivity.}

Figs. \ref{KA_C_dmin_Pt10} and \ref{KA_C_dmax_Pt10} present the results at the highest $P_t$ for the minimum and maximum Tx-Rx distances, respectively. The effect of brightness temperature and fading conditions remains significant and leads to decreases in ergodic capacity from 60 Mbps to 44.5 Mbps, as shown in Fig. \ref{KA_C_dmin_Pt10}, and from 32 Mbps to 17.5 Mbps, as shown in Fig. \ref{KA_C_dmax_Pt10}. However, increasing $P_t$ raises the ergodic capacity by about 16 Mbps under all conditions in Fig. \ref{KA_C_dmin_Pt10} compared to Fig. \ref{KA_C_dmin_Pt1}. A similar improvement is observed between Fig. \ref{KA_C_dmax_Pt10} and Fig. \ref{KA_C_dmax_Pt1}. \textit{This indicates that increasing the transmit power is an effective way to mitigate the effects of the $\lambda$-value and distance while achieving more stable performance.}
%%__%%__%%__%%__%%%%__%%__%%__%%__%%%%__%%__%%__%%__%%%%__%%__%%__%%__%%
%%__%%__%%__%%__%%%%__%%__%%__%%__%%%%__%%__%%__%%__%%%%__%%__%%__%%__%%
%%__%%__%%__%%__%%%%__%%__%%%%___Ka band___%%%%__%%__%%__%%__%%%%__%%__%
%%__%%__%%__%%__%%%%__%%__%%__%%__%%%%__%%__%%__%%__%%%%__%%__%%__%%__%%
%%__%%__%%__%%__%%%%__%%__%%__%%__%%%%__%%__%%__%%__%%%%__%%__%%__%%__%%
\begin{figure*}[!b]
\centering
\begin{subfigure}{\columnwidth}
\centering
\includegraphics[width=\linewidth]{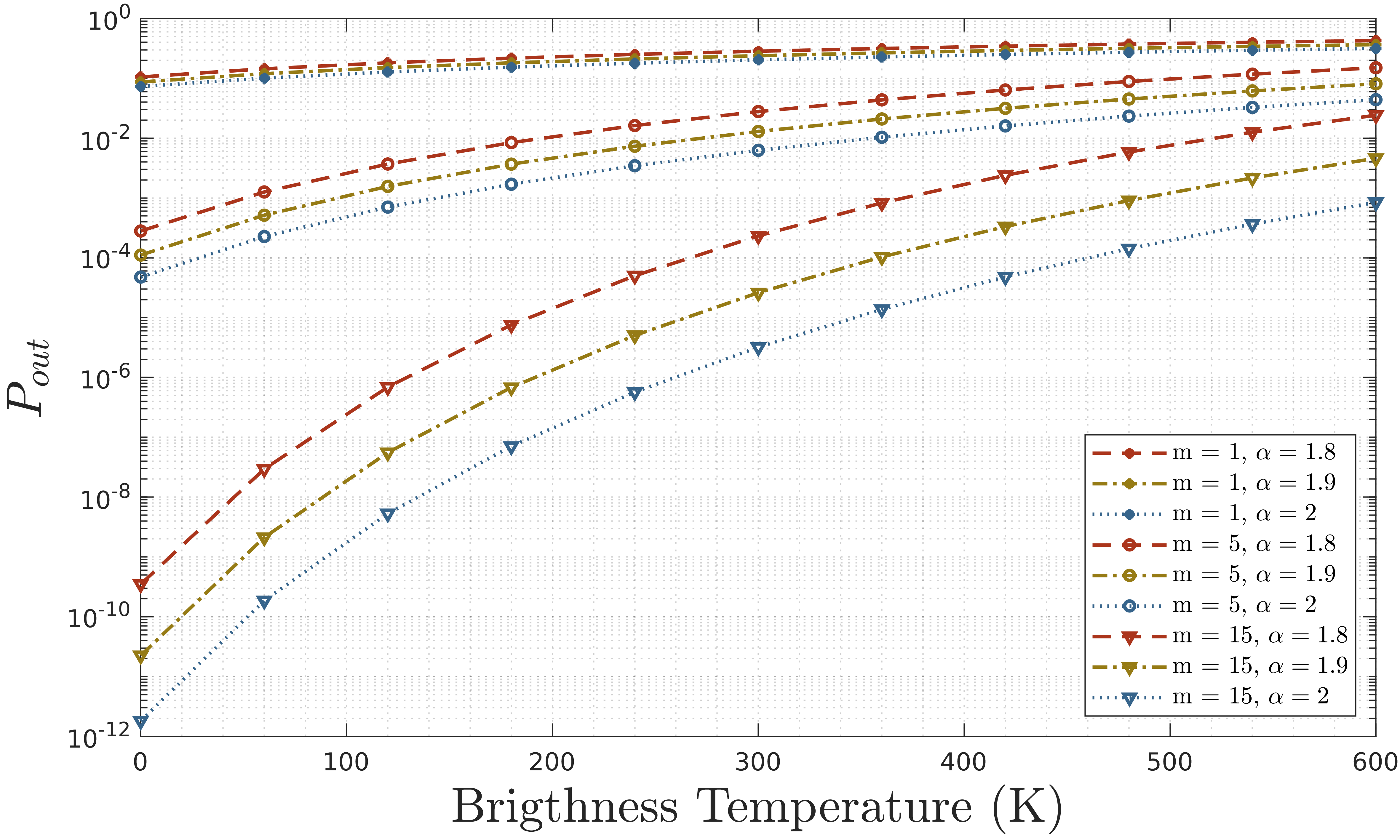}
\caption{$\gamma_{th}=15$ dB, $d = 10\times10^6$ m, $P_t = 1$ W.}
\label{KA_OUTAGE_dmin_Pt1}
\end{subfigure}\hspace*{\fill}
\begin{subfigure}{\columnwidth}
\centering
\includegraphics[width=\linewidth]{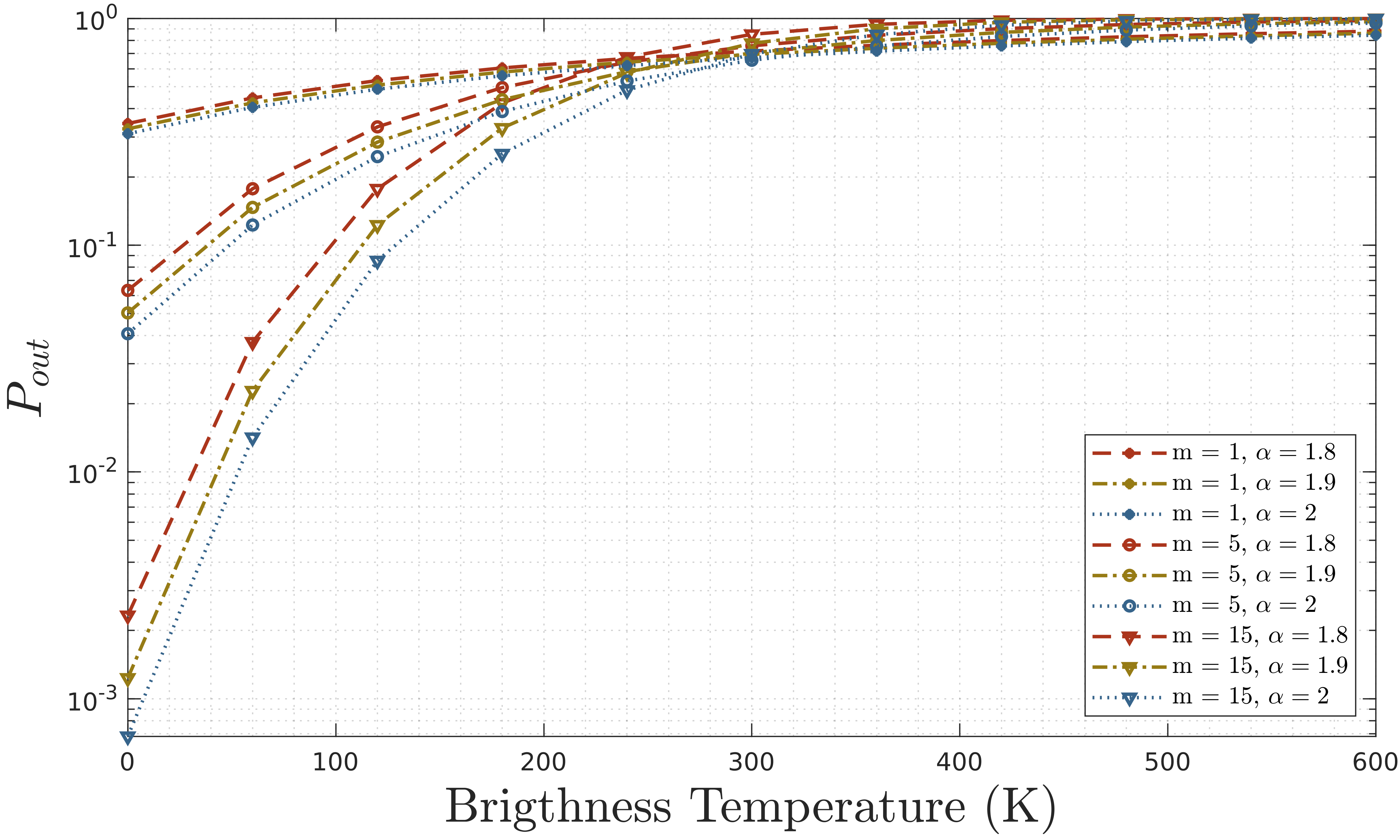}
\caption{$\gamma_{th}=5$ dB, $d = 70\times10^6$ m, $P_t = 1$ W.}
\label{KA_OUTAGE_DMAX_Pt1}
\end{subfigure}\hspace*{\fill}
\\
\begin{subfigure}{\columnwidth}
\centering
\includegraphics[width=\linewidth]{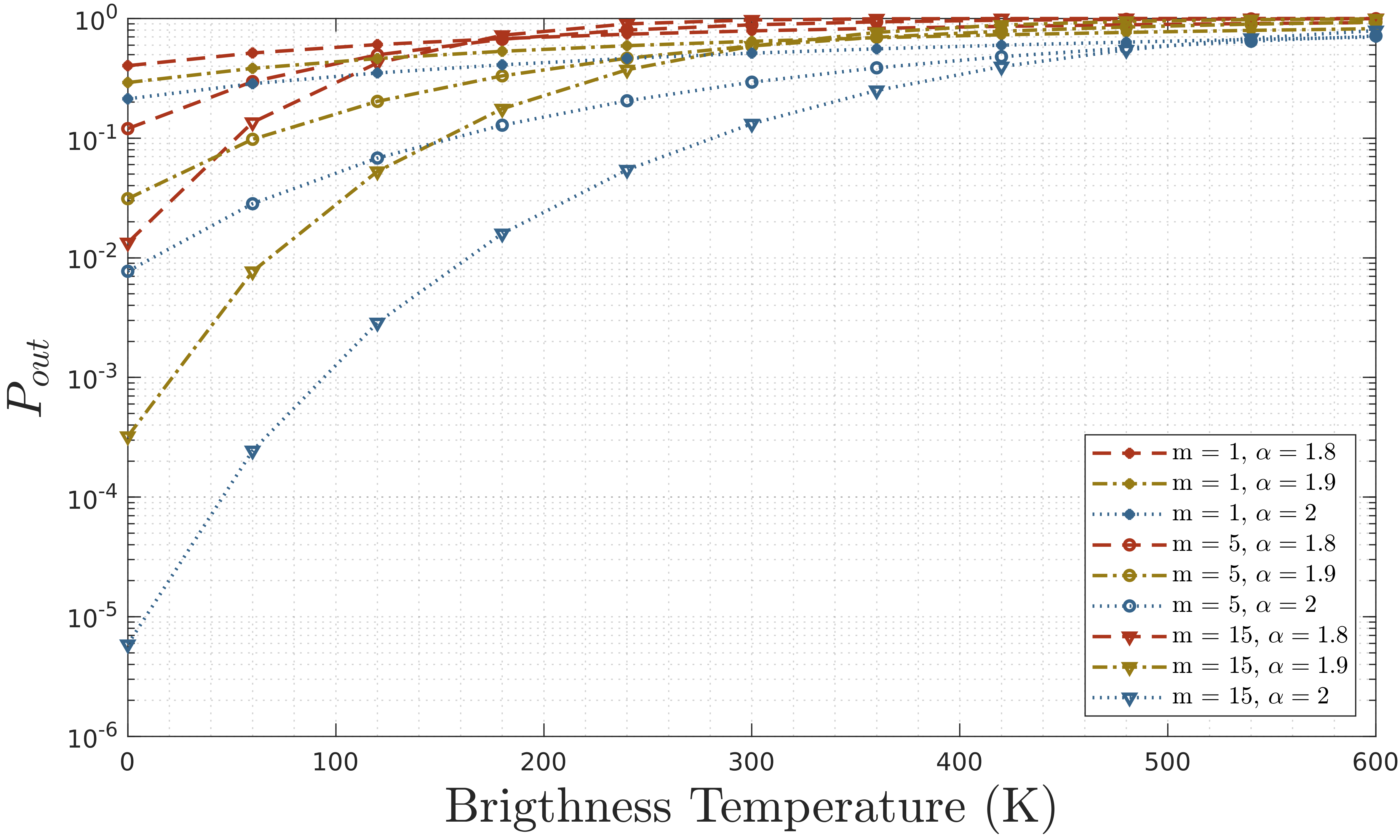}
\caption{$\gamma_{th}=30$ dB, $d = 10\times10^6$ m, $P_t = 10$ W.}
\label{KA_OUTAGE_dmin_Pt10}
\end{subfigure}\hspace*{\fill}
\begin{subfigure}{\columnwidth}
\centering
\includegraphics[width=\linewidth]{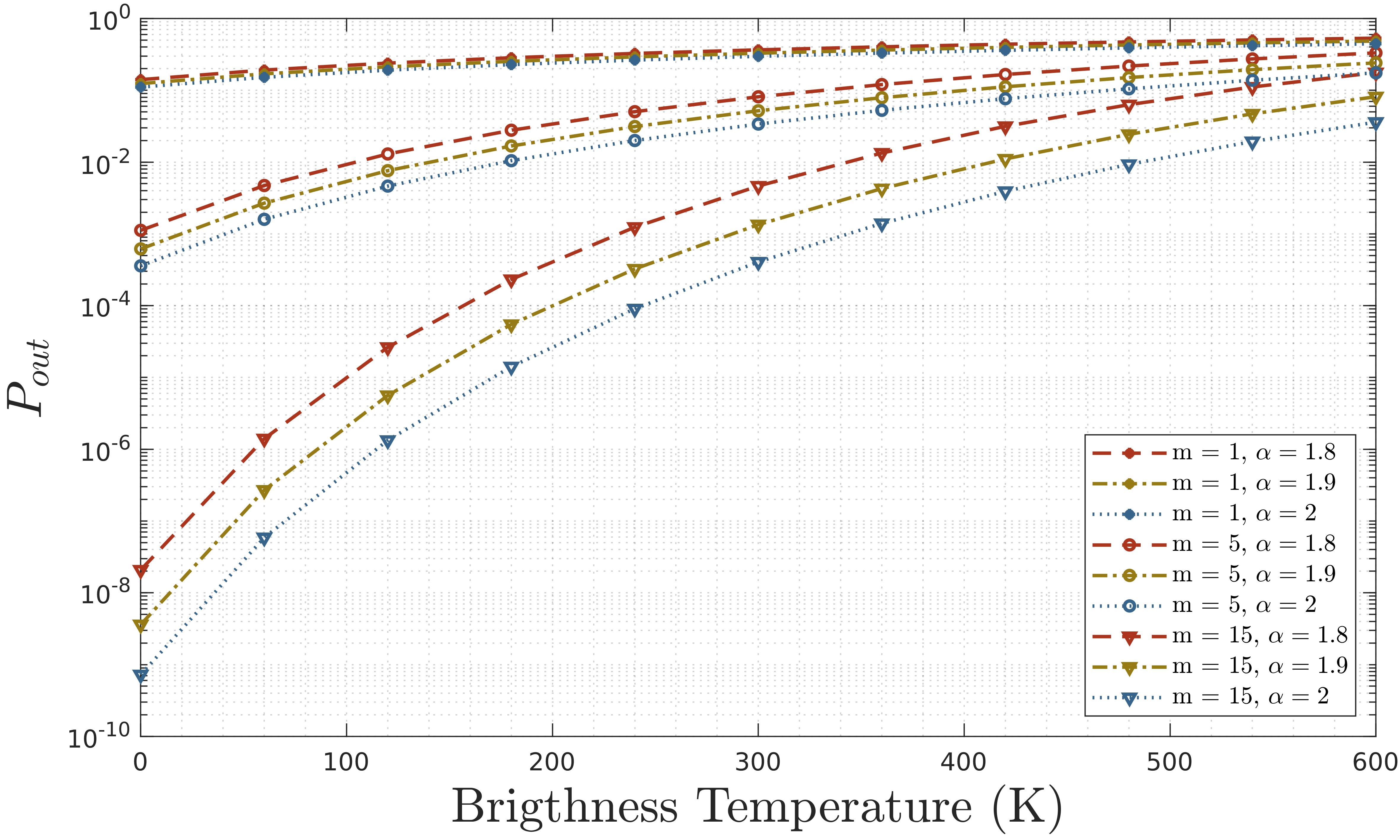}
\caption{$\gamma_{th}=10$ dB, $d = 70\times10^6$ m, $P_t = 10$ W.}
\label{KA_OUTAGE_DMAX_Pt10}
\end{subfigure}
\caption{The comparison of the outage probability for the Ka bandwidth the upper bound in equation (\ref{Pout_nakagami_definite}) depending on the brightness temperature and under different system settings.}
\label{KA_OUTAGE}
\end{figure*}
%%__%%__%%__%%__%%
%%__%%__%%__%%__%%
%%__%%__%%__%%__%%
%%__%%__%%__%%__%%

%%__%%__%%__%%__%%%%__%%__%%_______S BAND CAPACITY_____%%%%__%%__%%__%%__%%%%__%%__%%__%%__%%
Fig. \ref{S_C} illustrates the relationship between the brightness temperature and the lower bound of the ergodic capacity for different system configurations in the S band. The results show that the ergodic capacity ranges between 70 Kbps and 4.1 Mbps, depending on the system and channel parameters. Fig. \ref{S_C_dmin_Pt1} shows the results for the minimum Tx-Rx distance and the lowest transmit power. The increase in brightness temperature and the deteriorating fading conditions lead to a decrease in the ergodic capacity from 2.45 Mbps to 1.15 Mbps. In fact, communication signals in S band are less vulnerable to noise fluctuations than the Ka band due to its longer wavelength and thus the higher $P_r$. However, unlike the results shown in Fig. \ref{KA_C_dmin_Pt1} for the Ka band, the effect of noise stability on the ergodic capacity is clearly visible due to the small range of the y-axis. \textit{This also suggests that accurate estimation of the $\lambda$ and $\alpha$ values in S band is essential to ensure lower susceptibility to performance degradation in average capacity.} 

Fig. \ref{S_C_dmax_Pt1} shows the results for the maximum distance with the lowest $P_t$. Since the distance between the Lunar Gateway and the Moon is the greatest, the lower bounds of the ergodic capacity fall in the range of 70 to 380 Kbps. It also clearly shows how the $\alpha$-values together with the brightness temperature affect the ergodic capacity. The gap between the worst fading conditions significantly narrows compared to the other results in Figs. \ref{KA_C} and \ref{S_C}. Similar to Fig. \ref{KA_C}, we observe that capacity is higher in all parameter settings in Fig. \ref{S_C_dmin_Pt10}, which presents the results for the minimum Tx-Rx distance with the highest $P_t$. The average capacity decreases from 4.05 Mbps to 2.6 Mbps with increasing brightness temperature and worse fading conditions. At the maximum Tx-Rx distance and the highest $P_t$, as shown in Fig. \ref{S_C_dmax_Pt10}, the lower bounds of the ergodic capacity fall in the range of 0.4 Mbps to 1.4 Mbps due to larger path loss caused by the longer distance. Despite the increase in $P_t$, the influence of the noise parameters remains evident in Fig. \ref{S_C_dmax_Pt10}, as it is in Figs. \ref{S_C_dmin_Pt1} and \ref{S_C_dmax_Pt1}.
%%%%%%%%%%%%%%%%%%%%
%%%%%%%%%%%%%%%%%%%%
%%%%%%%%%%%%%%%%%%%%
%%%%%%%%%%%%%%%%%%%%
%%%%%%%%%%%%%%%%%%%%
%%%%%%%%%%%%%%%%%%%%
%%%%%%%%%%%%%%%%%%%%
%%%%%%%%%%%%%%%%%%%%
%%%%%%%%%%%%%%%%%%%%
%%%%%%%%%%%%%%%%%%%%
%%%%%%%%%%%%%%%%%%%%
%%%%%%%%%%%%%%%%%%%%
\subsection{Outage Probability Upper Bound}
\begin{figure*}[!ht]
\centering
\begin{subfigure}{\columnwidth}
\centering
\includegraphics[width=\linewidth]{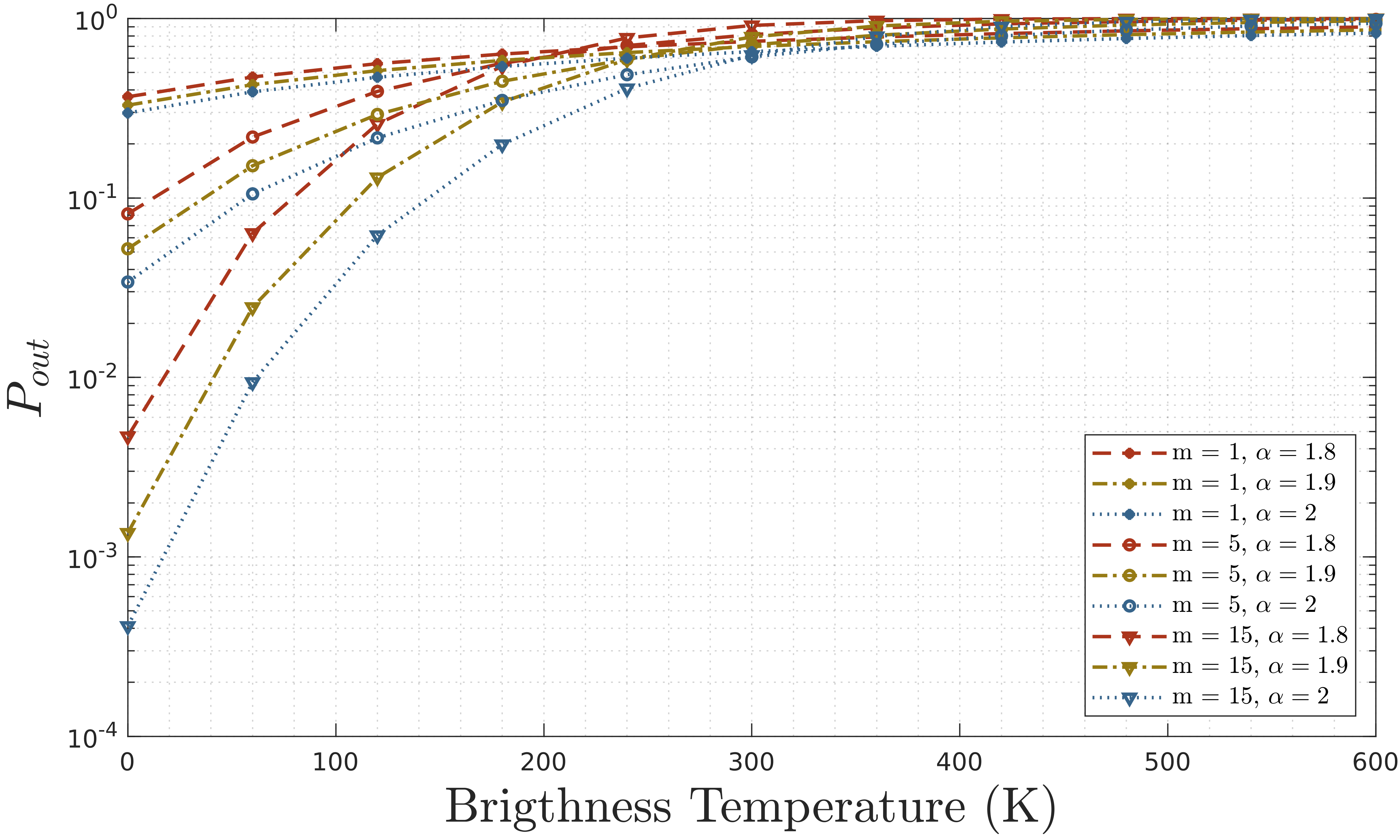}
\caption{$\gamma_{th}=10$ dB, $d = 10\times10^6$ m, $P_t = 1$ W.}
\label{S_OUTAGE_dmin_Pt1}
\end{subfigure}\hspace*{\fill}
\begin{subfigure}{\columnwidth}
\centering
\includegraphics[width=\linewidth]{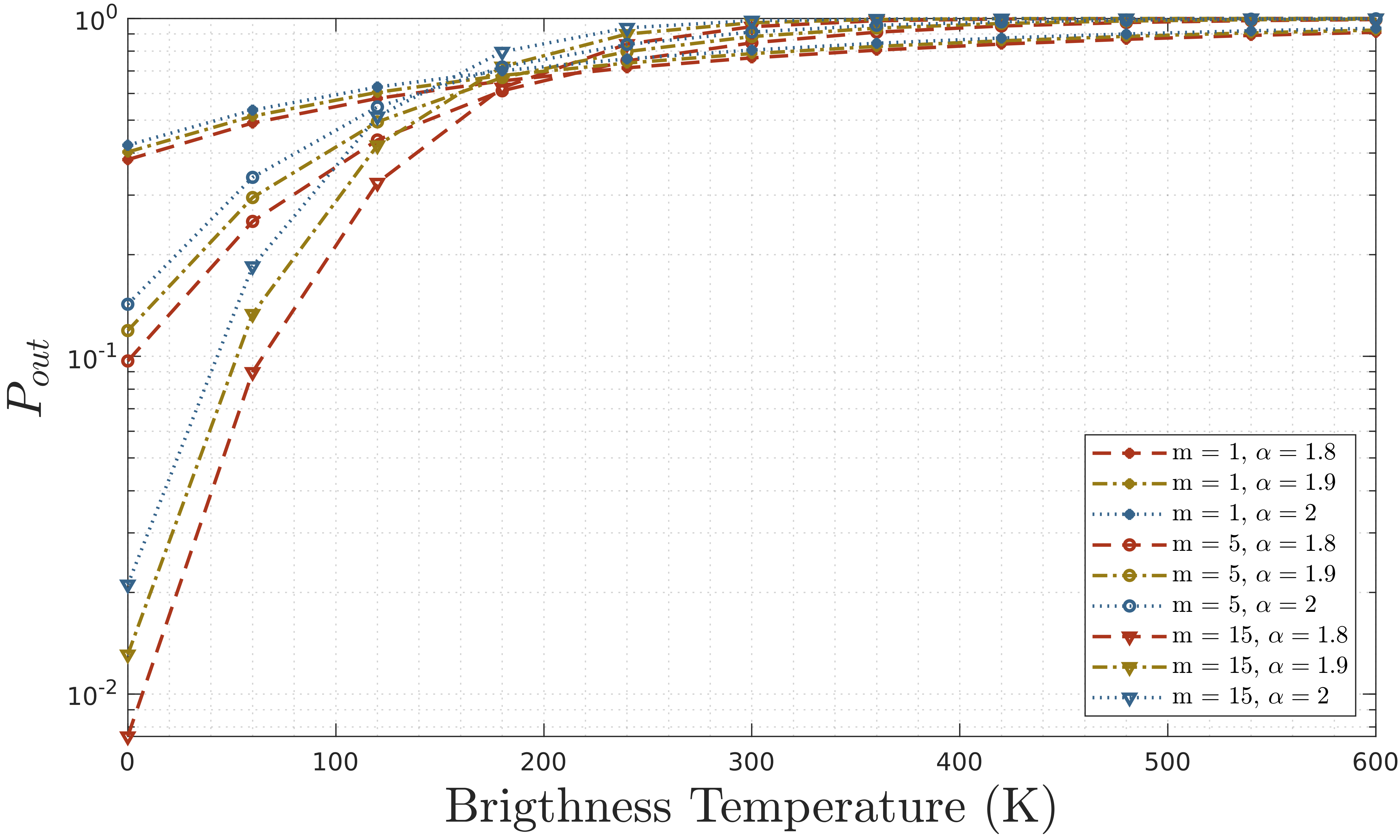}
\caption{$\gamma_{th}=-5$ dB, $d = 70\times10^6$ m, $P_t = 1$ W.}
\label{S_OUTAGE_DMAX_Pt1}
\end{subfigure}\hspace*{\fill}
\\
\begin{subfigure}{\columnwidth}
\centering
\includegraphics[width=\linewidth]{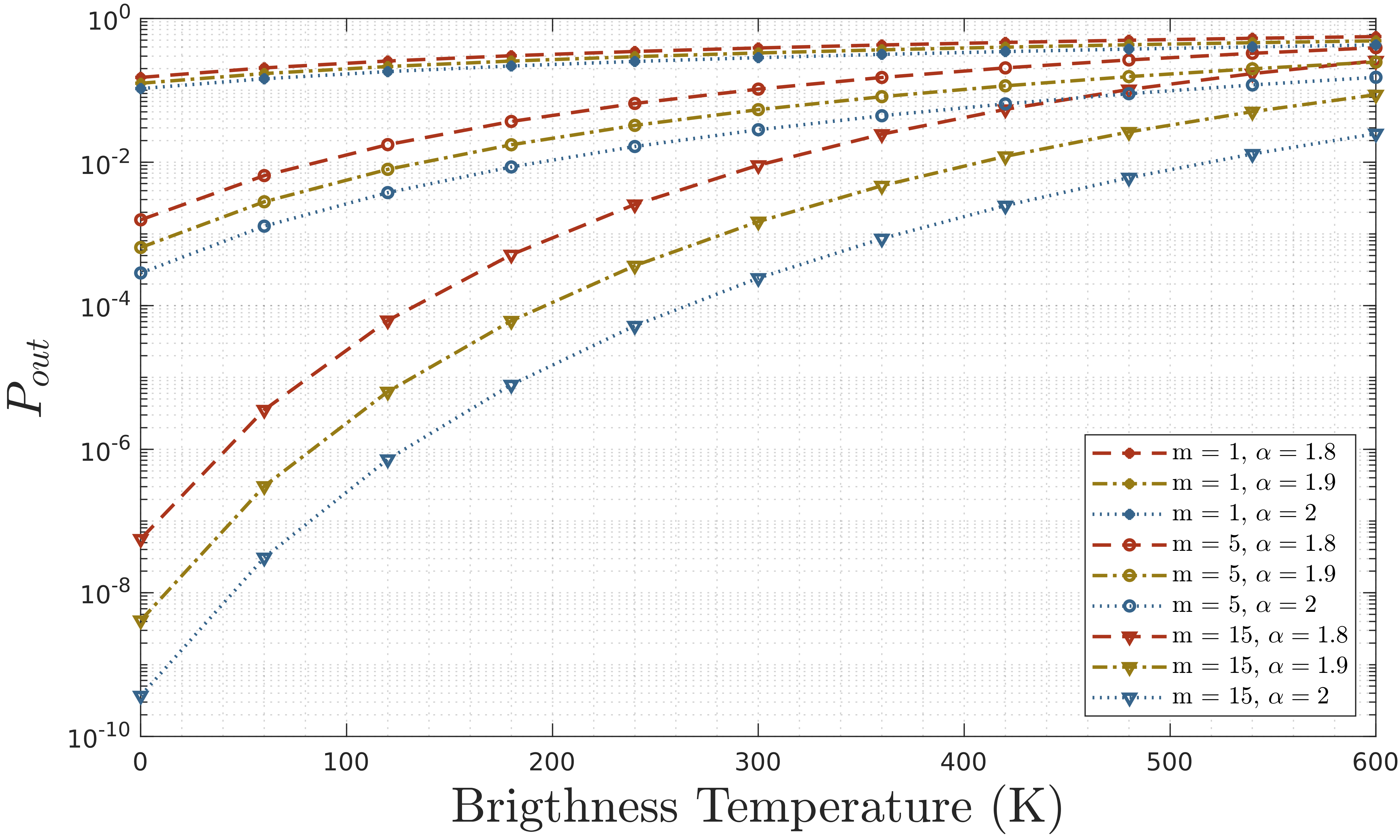}
\caption{$\gamma_{th}=15$ dB$, d = 10\times10^6$ m, $P_t = 10$ W.}
\label{S_OUTAGE_dmin_Pt10}
\end{subfigure}\hspace*{\fill}
\begin{subfigure}{\columnwidth}
\centering
\includegraphics[width=\linewidth]{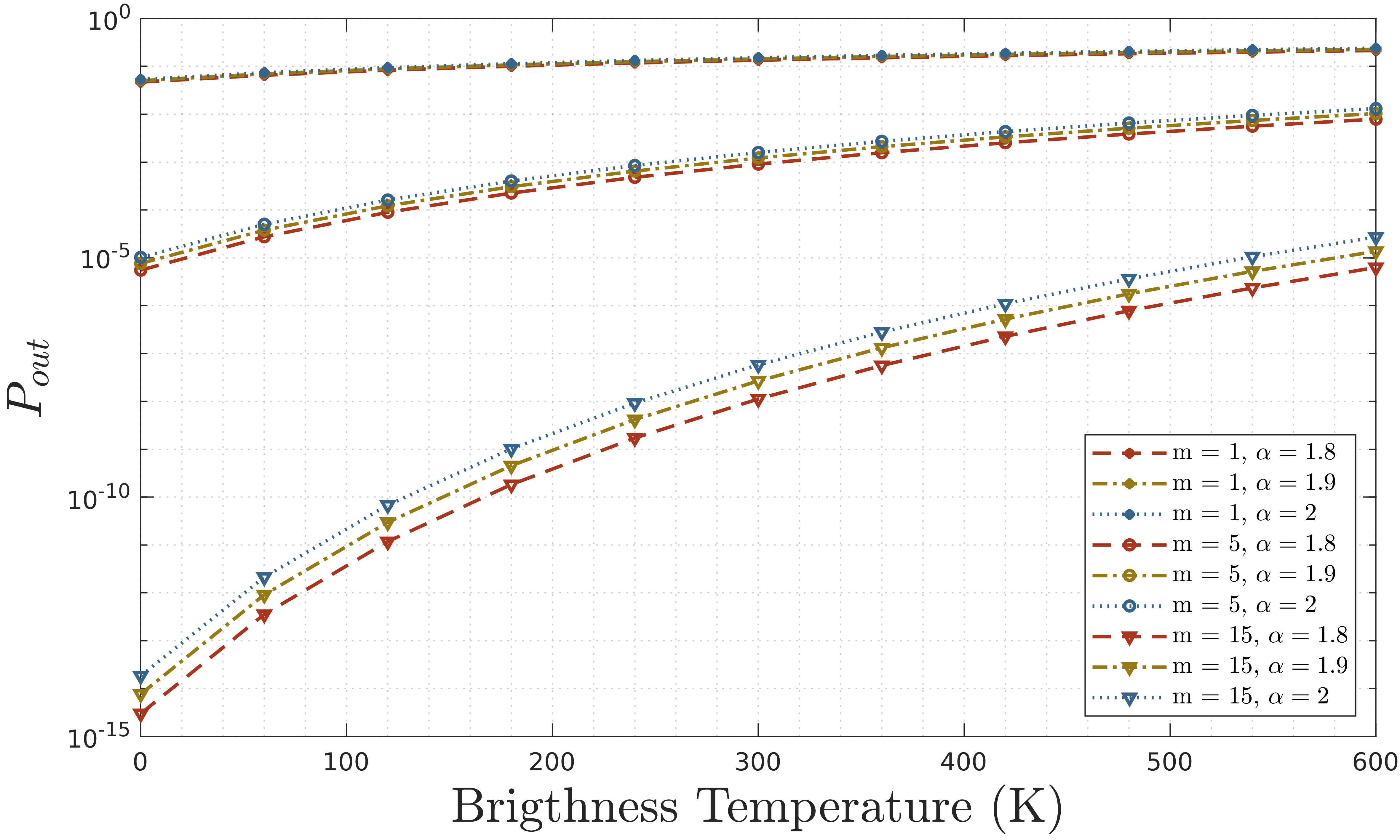}
\caption{$\gamma_{th}=-5$ dB, $d = 70\times10^6$ m, $P_t = 10$ W.}
\label{S_OUTAGE_DMAX_Pt10}
\end{subfigure}\hspace*{\fill}
\caption{The comparison of the outage probability for the S bandwidth the upper bound in equation (\ref{Pout_nakagami_definite}) depending on the brightness temperature and under different system settings.}
\label{S_OUTAGE}
\end{figure*}

In the previous section, we presented the results for ergodic capacity under different system configurations. In particular, we observed how the noise scale parameter varies significantly while the noise stability index remains constant due to the brightness temperature. We have addressed how the ergodic capacity becomes more vulnerable depending on the noise scale parameter if the $\alpha$-value is not accurately estimated. In this section, we propose to focus on the instantaneous capacity performance over the outage probability. This approach allows us to directly observe the impact of the $\alpha$-value on the capacity. The results based on equation (\ref{Pout_nakagami_definite}) are presented separately for Ka and S bands, along with comprehensive analyzes in Figs. \ref{KA_OUTAGE} and \ref{S_OUTAGE}.
%%%%%%%%%%%%%%%%%%%%%%%%%%%%%___S*BAND___%%%%%%%%%%%%%%%%%%%%%%%%%%%%%
%%__%%__%%__%%__%%
%%__%%__%%__%%__%%
%%__%%__%%__%%__%%
%%__%%__%%__%%__%%

Fig. \ref{KA_OUTAGE_dmin_Pt1} illustrates the outage probability at the minimum Tx-Rx distance and lowest transmit power. The differences in the outage probabilities under different fading conditions can be clearly seen. Notably, the outage probability for $m=1$, which corresponds to Rayleigh fading, is intolerable for reliable cislunar communication. As expected, the outage probability increases together with brightness temperature even under more favorable fading conditions. More importantly, the figure shows how variations in $\alpha$-values can lead to unexpected communication interruptions for the Ka band. For example, $P_\textup{{out}}$ at $T_B=120$ K and $m=15$ is less than $10^{-8}$ for $\alpha=2$, but it increases to about $10^{-6}$ for $\alpha=1.8$.

As the distance between transmitter and receiver reaches its maximum at the lowest $P_t$, $P_\textup{{out}}$ increases significantly, as can be seen in Fig. \ref{KA_OUTAGE_DMAX_Pt1}. Although the threshold value for the instantaneous SNR is lowered by 10 dB compared to Fig. \ref{KA_OUTAGE_dmin_Pt1}, the outage probabilities are unacceptable at brightness temperatures greater than 100 K, even in the case of $m=15$. This is especially a critical problem for real-time control applications of CSNs and missions. Fig. \ref{KA_OUTAGE_DMAX_Pt1} also illustrates the joint influence of the propagation and noise stability conditions as well as the brightness temperature. However, it also shows that the considerations outlined in this article are crucial to ensure reliable and dynamic CSNs.

Fig. \ref{KA_OUTAGE_dmin_Pt10} compares the $P_\textup{{out}}$ at the minimum distance and the highest transmit power. The effect of the noise stability index becomes clearer compared to the other figures when the threshold value for the instantaneous SNR — i.e. the target data rate — is increased. This shows that the impulsiveness of the communication signal leads to more communication interruptions for missions requiring high data rates. \textit{It also sheds light on the trade-offs between mission design and objectives. For example, a mission operating in a more challenging propagation environment may require a power system that ensures communication at a higher transmit power, or alternatively be designed with a lower data rate objective.} Fig. \ref{KA_OUTAGE_DMAX_Pt10} shows the results at the maximum distance and highest $P_t$. Decreasing the threshold for the received instantaneous SNR does not prevent the increase in the outage probability, which is primarily caused by the increasing path loss. \textit{This suggest that higher transmit power is required to maintain reliable cislunar communication, especially at increasing Tx-Rx distance and under worse fading conditions.}

%%%%%___S--band__outage_results_____%%%%
Fig. \ref{S_OUTAGE} compares the outage probabilities for the S-band in accordance with the system set-up and configurations in Section \ref{results}-\ref{sim_setup}. Fig. \ref{S_OUTAGE_dmin_Pt1} shows the results at minimum distance and lowest transmit power. The results are acceptable except for the worst fading conditions. However, the risk for communication failures is generally unacceptable in Fig. \ref{S_OUTAGE_DMAX_Pt1}, which presents the results at the maximum distance and the lowest transmit power. The outage probability is more sensitive to the noise scale parameter than to the noise stability index as the brightness temperature increases, similar to the results observed for the ergodic capacity. However, this effect is not seen in Fig. \ref{KA_OUTAGE_DMAX_Pt1}, which shows the results for the same system configurations in Ka band. The reason for this difference is that the S band has a larger wavelength, resulting in lower received power under the same system conditions. \textit{This indicates that possible solutions to improve $P_r$ should be considered when developing communication systems, such as reducing system losses or increasing transmit power.}

Fig. \ref{S_OUTAGE_dmin_Pt1} presents the results for the $P_\textup{{out}}$ at the minimum distance and the lowest $P_t$ for an instantaneous SNR threshold of 15 dB. We observe that the performance, especially for $m=15$, improves for all values of the noise stability index and allows higher data rate targets. In contrast, for $m=1$, either the target data rates — $\gamma_{th}$ in other words — should be reduced or the transmit power should be increased to ensure reliable communication. Fig. \ref{S_OUTAGE_DMAX_Pt10} shows the $P_\textup{{out}}$ at maximum distance and highest $P_t$. If we compare the results in Figs. \ref{S_OUTAGE_dmin_Pt1} and \ref{S_OUTAGE_DMAX_Pt10}, increasing the transmit power improves the outage probability significantly. However, the instantaneous capacity remains very sensitive to the values for $\alpha$ and $\lambda$, as in Fig. \ref{S_OUTAGE_DMAX_Pt1}, due to the lower received power.

Overall, the results illustrate the significant influence of brightness temperature and variations in noise properties, but also show the importance of their accurate estimation. Furthermore, the effects of distances due to cislunar geometry, changing propagation and noise conditions, and power and bandwidth constraints become clear with the results. In this way, they highlight the trade-offs between designs, objectives and system constraints of missions through the extensive system and channel settings. These results can also serve as inspiration for further studies focusing on deep space communications, as Lunar Gateway is also intended to support deep space missions. However, deep space missions are likely to be more challenging in many aspects, as existing knowledge is still limited compared to cislunar space.
%%%%%%%%%%%%%%%%%%%%%%%%%%%%%%%%%%%%%%%%%%%%%%%%%%%%%%%%%%%%%%%
%%%%%%%%%%%%%%%%%%%%%%%%%%%%%%%%%%%%%%%%%%%%%%%%%%%%%%%%%%%%%%%
%%%%%%%%%%%%%%%%%%%%%__Conclusions__%%%%%%%%%%%%%%%%%%%%%%%%%%%
%%%%%%%%%%%%%%%%%%%%%%%%%%%%%%%%%%%%%%%%%%%%%%%%%%%%%%%%%%%%%%%
%%%%%%%%%%%%%%%%%%%%%%%%%%%%%%%%%%%%%%%%%%%%%%%%%%%%%%%%%%%%%%%
\section{CONCLUSIONS AND FURTHER WORK}\label{conclusion}
With respect to the emerging CSNs with outstanding goals, we propose system-level and theoretical analyzes for cislunar communication to give an outlook on their dynamic and robust designs in the future. Therefore, we first discussed the aspects of communication in cislunar space and the corresponding considerations, such as temperature fluctuation, hilly terrain with sunlit and shadowed areas, and high mobility. To analyze the reliable communication, we proposed the ergodic capacity and outage probability. Prior to our theoretical analyzes, we investigated the relationship between the temperature variations on the lunar surface and the Lunar Gateway receiver. Since this relationship and other aspects of cislunar communication can lead to impulsivity and different propagations of communication signals, we introduced a signal model that takes these aspects into account. Then, we performed our theoretical analysis. Furthermore, in our results, we extend the analyzes for both high and low data rate mission targets with comprehensive system configurations. The results clearly show the significant influence of many phenomena on the ergodic capacity and outage probability, but also to provide insights into the future communication systems of CSNs. For further studies, we plan to extend the design considerations in light of our current analysis to further develop CSNs.
%_____________________________________________________________%
%_____________________________________________________________%
%%%%%%%%%%%%%%%%%%%%%%%%%%%%%%%%%%%%%%%%%%%%%%%%%%%%%%%%%%%%%%%
%%%%%%%%%%%%%%%%%%%%%%%%%%%%%%%%%%%%%%%%%%%%%%%%%%%%%%%%%%%%%%%
\section*{ACKNOWLEDGMENT}
This work was supported in part by the Tier 1 Canada Research Chair program and the Natural Sciences and Engineering Research Council of Canada (NSERC) Discovery Grant program.
%%__%%__%%__%%__%%%%__%%__%%__%%__%%%%__%%__%%__%%__%%%%__%%__%%__%%__%%%%__%%__%%__%%__%%%%%%%%%%%%%%%%%%%%%%%%%%%%%%%%%%%%%%%%%%%%%%%%%%%%%%%%%%%%%%%%%%%%%%%%%%%%%%%%%%%%%%%%%%%%%%%
%%__%%__%%__%%__%%
%%__%%__%%__%%__%%
%%__%%__%%__%%__%%
%%__%%__%%__%%__%%
%%__%%__%%__%%__%%
%%__%%__%%__%%__%%
{\appendices
%%%%%%%%%%%%%%%%%%%%%%%%%%%%%%%%%%%%%%%%%%%%%%%%%%%%%%%%%%%%%%%
%%%%%%%%%%%%%%%%%%%%%%__Appendix A__%%%%%%%%%%%%%%%%%%%%%%%%%%%
%%%%%%%%%%%%%%%%%%%%%%%%%%%%%%%%%%%%%%%%%%%%%%%%%%%%%%%%%%%%%%%
%%%%%%%%%%%%%%%%%%%%%%%%%%%%%%%%%%%%%%%%%%%%%%%%%%%%%%%%%%%%%%%
\section*{APPENDIX A}
\label{appendixA}
Let define $n =n_1 +jn_2$ where $n_i = \sqrt{A_i}G_i$ for $i=1,2$. Here, ${A_i} \sim S(\alpha /2,1,[\cos (\pi \alpha /4)]^{2/\alpha },0)$ and ${G_i} \sim \mathcal{N}(0,\sigma ^{2})$. Since the alpha-stable distribution is Gaussian when $\alpha=2$ and $\beta=0$, $G_i \sim S \left ( 2, 0,\frac{\sigma}{\sqrt{2}},0 \right)$ is defined as well. Then, the characteristic function of $n_i$ is written as
\begin{equation}
\begin{split}
\Phi_{n_i} (t) &=  \mathbb {E}\left[ \exp(jtn_i))\right]      \\
&=  \mathbb {E}\left[ \exp\left(jt{A_i^{1/2}}G_i)\right) \right]       \\ 
&=  \mathbb {E}\left[ 
\left.\begin{matrix} \mathbb {E}\left[\exp \left(-j \left|t\right|^2  
\left(\frac{\sigma}{\sqrt{2}}\right)^2    A_i \right) \right]  \end{matrix}\right|
A_i          \right].                                                     \\ 
%  &=
\end{split}
\end{equation}
By substituting the Laplace transform of $A_i$, $\mathbb {E}\left[\exp\left(-sA_i\right)\right] = \exp\left(-s^{(\alpha/2)})\right)$ 
\begin{equation}
\Phi_{n_i} (t) = \mathbb {E}\left[\exp\left(-j\left|t\right|^\alpha 
\left( \frac{\sigma}{\sqrt{2}} \right)^\alpha 
\right)\right]. 
\end{equation}
We show that $n_i \sim S \left(\alpha, 0,\frac{\sigma}{\sqrt{2}},0 \right)$, $n$ is the sum of two $\text{S}\alpha\text{S}$ variables in other words. Then, it can be seen that $n \sim S \left(\alpha ,0,2^{\left ( \frac{1}{\alpha}-\frac{1}{2} \right )}\sigma,0 \right)$ from Property 4.
%%%%%%%%%%%%%%%%%%%%%%%%%%%%%%%%%%%%%%%%%%%%%%%%%%%%%%%%%%%%%%%
%%%%%%%%%%%%%%%%%%%%%%__Appendix B__%%%%%%%%%%%%%%%%%%%%%%%%%%%
%%%%%%%%%%%%%%%%%%%%%%%%%%%%%%%%%%%%%%%%%%%%%%%%%%%%%%%%%%%%%%%
%%%%%%%%%%%%%%%%%%%%%%%%%%%%%%%%%%%%%%%%%%%%%%%%%%%%%%%%%%%%%%%
\section*{APPENDIX B}\label{appendixB}
Let $Y= \left | h \right |^\alpha$ and the cdf of $Y$ is calculated as 
\begin{equation}
\begin{split}
F_{Y}(y) &= P(\left | h \right |^{\alpha}\leq y) \\ 
         &= P(\left | h \right |\leq y^{1/\alpha}) \\
         &= F_{\left | h \right |}(y^{1/\alpha}).
\end{split}
\end{equation}
By using above substitution, the pdf of $\left | h \right |^\alpha$ is derived as
\begin{equation}
\begin{split}
f_{ \left | h \right |^\alpha}(y) 
&=  \frac{\partial F_{\left | h \right |}(y^{1/\alpha}) }{\partial y}  \\ 
&=   \frac{1}{\alpha}f_{\left | h \right |}(y^{1/\alpha})y^{\frac{1}{\alpha}-1}  
\end{split}
\end{equation}
$f_{ \left | h \right |^\alpha}(y)$ is obtained by inserting the pdf of Nakagami-m distribution as follows
\begin{equation}
\label{pdf_halpha}
%f_{ \left | h \right |^\alpha}(y)
     f_{ \left | h \right |^\alpha}(y) =  \frac{2m^m y^{-1}}{\Omega \alpha \Gamma(m)}  \exp\left[-\frac{m}{\Omega} y^{\frac{2}{\alpha}}  \right]  y^{\frac{2m}{\alpha}} .
\end{equation}
By employing a change of variable $\left | h \right |^\alpha = \frac{\gamma \xi }{\bar{\gamma}}$ within the cdf function of the instantaneous SNR, we derived the cdf of instantaneous SNR as a function of $F_{ \left | h \right |^\alpha}(y)$ below.
\begin{equation}
\label{snrcdf}
\begin{split}
F_{{\gamma}}(\gamma) 
%&=  P({\gamma} \leq \gamma)  \\ 
&=  P(\left | h \right |^{\alpha} \frac{\bar{\gamma}}{\xi}  \leq \gamma)\\
&= P(\left | h \right |^{\alpha}  \leq \gamma \frac{\xi}{\bar{\gamma}} ) \\ 
&= F_{ \left | h \right |^\alpha}(\gamma \frac{\xi}{\bar{\gamma}})
\end{split}
\end{equation}
Using equation (\ref{snrcdf}), $f_{\gamma}(\gamma) $ is derived as follows
\begin{equation}
\begin{split}
f_{\gamma}(\gamma)  
&= \frac{\xi}{\bar{\gamma}} f_{ \left | h \right |^\alpha}(\gamma \frac{\xi}{\bar{\gamma}})\\
&=  \frac{2m^m \gamma^{-1}}{ \Omega \alpha \Gamma(m)}     \exp\left[-\frac{m}{\Omega}     \left( \frac{\gamma\xi}{\bar{\gamma}} \right)^{\frac{2}{\alpha}}\right]  \left( \frac{\gamma \xi }{\bar{\gamma}} \right)^{\frac{2m}{\alpha}}.
\end{split}
\end{equation}
% %%%%%%%%%%%%%%%%%%%%%%%%%%%%%%%%%%%%%%%%%%%%%%%%%%%%%%%%%%%%%%%
% %%%%%%%%%%%%%%%%%%%%%%__Appendix C__%%%%%%%%%%%%%%%%%%%%%%%%%%%
% %%%%%%%%%%%%%%%%%%%%%%%%%%%%%%%%%%%%%%%%%%%%%%%%%%%%%%%%%%%%%%%
% %%%%%%%%%%%%%%%%%%%%%%%%%%%%%%%%%%%%%%%%%%%%%%%%%%%%%%%%%%%%%%%
\section*{APPENDIX C}\label{appendixC}
The average capacity for a Nakagami fading channel is lower bounded by substituting (\ref{Naksnr_pdf}) into (\ref{gold_cap}) as follows%the well-known ergodic capacity formula as follows
\begin{equation}\label{nak_c1}
\begin{split}
\overline{C} \geq 
& \int_{0}^{\infty}  \frac{2m^m \gamma^{-1}}{ \Omega \alpha \Gamma(m)} \log_{2}(1+\gamma) \\
&\times   \exp\left[-\frac{m}{\Omega}     \left( \frac{\gamma\xi}{\bar{\gamma}} \right)^{\frac{2}{\alpha}}\right]  \left( \frac{\gamma \xi }{\bar{\gamma}} \right)^{\frac{2m}{\alpha}}     d\gamma
\end{split}
\end{equation}
The above integration is manipulated by utilizing substitutions in equations (\ref{meijer2}) and (\ref{meijer1}).
\begin{equation}
\label{meijer2}
    G_{2,2}^{1,2}\left [ \gamma \left| \begin{matrix} 1,1  \\  1,0  \end{matrix} \right.\right ] =\ln{(1+\gamma)}.  
\end{equation}
\begin{equation}
\label{meijer1}
G_{0,1}^{1,0} \left [  \frac{  \left ( \gamma\xi /\bar{\gamma} \right )^\frac{2}{\alpha} }{\Omega m^{\text{-1}}} \left| \begin{matrix} {\text{-}}  \\  m  \end{matrix} \right.\right ]    =
\frac{  \left ( \gamma\xi /\bar{\gamma} \right )^\frac{2m}{\alpha} }{\Omega^m m^{\text{-m}}}
\exp\left [-\frac{  \left ( \gamma\xi /\bar{\gamma} \right )^\frac{2}{\alpha} }{\Omega m^{\text{-1}}}  \right ].
\end{equation}
Then, the equation (\ref{nak_c1}) becomes as
\begin{equation}
\begin{split}
\overline{C} &\geq \frac{2 \Omega^{m-1}}{\alpha^2 \ln{(2)} \Gamma(m)}  \\  
&\times \int_{0}^{\infty}{\gamma}^{\text{-}1}     G_{0,1}^{1,0} \left [  \frac{  \left ( \gamma\xi /\bar{\gamma} \right )^\frac{2}{\alpha} }{\Omega m^{\text{-1}}} \left| \begin{matrix} {\text{-}}  \\  m  \end{matrix} \right.\right ]     G_{2,2}^{1,2}\left [ \gamma \left| \begin{matrix} 1,1  \\  1,0  \end{matrix} \right.\right ]    d\gamma.
\end{split}
\end{equation}
The above integration is solved with the help of the equation in \cite{wolfram_meijerg,mellin} and the lower bound of the ergodic capacity for a Nakagami fading channel is derived as 
\begin{equation}
\begin{split}
\overline{C} 
&\geq \frac{ \Omega^{m-1} l}{ 2 \mathrm{ln}(2)  \Gamma(m)} \sqrt{\frac{k^{2m-3}}{(2\pi)^{2l+k-3}}}   \\
&\times G_{2l,k+2l}^{k+2l,l} \left[ 
\frac{\left( {\xi } / \bar{\gamma}  \right)^{l}}{\left( k \Omega m^\text{-1}  \right)^k} 
\left| \begin{matrix} I(l,0),I(l,1) \\  I(k,m),I(l,0),I(l,0)  \end{matrix}  \right.\right] .
\end{split}
\end{equation}
%%%%%%%%%%%%%%%%%%%%%%%%%%%%%%%%%%%%%%%%%%%%%%%%%%%%%%%%%%%%%%%
%%%%%%%%%%%%%%%%%%%%%%__Appendix D__%%%%%%%%%%%%%%%%%%%%%%%%%%%
%%%%%%%%%%%%%%%%%%%%%%%%%%%%%%%%%%%%%%%%%%%%%%%%%%%%%%%%%%%%%%%
%%%%%%%%%%%%%%%%%%%%%%%%%%%%%%%%%%%%%%%%%%%%%%%%%%%%%%%%%%%%%%%
\section*{APPENDIX D}\label{appendixD}
The upper bound of the outage probability is calculated by utilizing the equations (\ref{Naksnr_pdf}) and (\ref{outage_general}) as 
\begin{equation}
\label{appendixd1}
\begin{split}
P_{\textup{out}}  (\gamma) 
&\leq   \int  \frac{2m^m \xi }{ \Omega \alpha \Gamma(m) \bar{\gamma}}  \\  
&\times\exp\left[-\frac{m}{\Omega}     \left( \frac{\gamma\xi}{\bar{\gamma}} \right)^{\frac{2}{\alpha}}\right]  \left( \frac{\gamma \xi }{\bar{\gamma}} \right)^{\frac{2m}{\alpha}-1} d\gamma.
\end{split}
\end{equation}
The equation (\ref{appendixd1}) is written by defining the variable $u=\left( \gamma\xi /\bar{\gamma} \right )^\frac{2}{\alpha}$ as follows
\begin{equation}
P_{\textup{out}}  (\gamma) \leq   \int   \frac{m^m}{ \Omega \Gamma(m)}     \exp\left[-\frac{m}{\Omega} u\right] u^{m-1} du.
\end{equation}
Then, it is written with another change of variable $v=u^m$ as 
\begin{equation}
\begin{split}
P_{\textup{out}}  (\gamma) 
&\leq   \int   \frac{m^{m-1}}{ \Omega \Gamma(m)}   \exp\left[-\frac{m}{\Omega} v^{1/m} \right] dv \\
&\leq   %\left.
-\frac{\Omega^{m-1}  \Gamma\left(m, \frac{m}{\Omega}v^{1/m}    \right) }{\Gamma \left ( m \right )},  
%\right| 
\end{split}
\end{equation}
where the integration can be derived by using the equation in \cite{wolfram_incomplete}. By inverting variables, we obtained the upper bound of the outage probability as follows
\begin{equation}
P_{\textup{out}} (\gamma) \leq \Omega^{m-1}-\frac{\Omega^{m-1} \Gamma \left( m,\frac{  \left ( \gamma\xi /\bar{\gamma} \right )^\frac{2}{\alpha} }{\Omega m^{\text{-1}}} \right) }{\Gamma \left ( m \right )}.
\end{equation}
%%%%%%%%%%%%%%%%%%%%%%%%%%%%%%%%%%%%%%%%%%%%%%%%%%%%%%%%%%%%%%%
}
%%__%%__%%__%%__%%
%%__%%__%%__%%__%%
%%__%%__%%__%%__%%
%\bibliographystyle{unsrt}
\bibliographystyle{IEEEtran}
\bibliography{GC_2024}
%%__%%__%%__%%__%%
%%__%%__%%__%%__%%
%%__%%__%%__%%__%%

\begin{IEEEbiography}[{\includegraphics[width=1in,height=1.25in,clip,keepaspectratio]{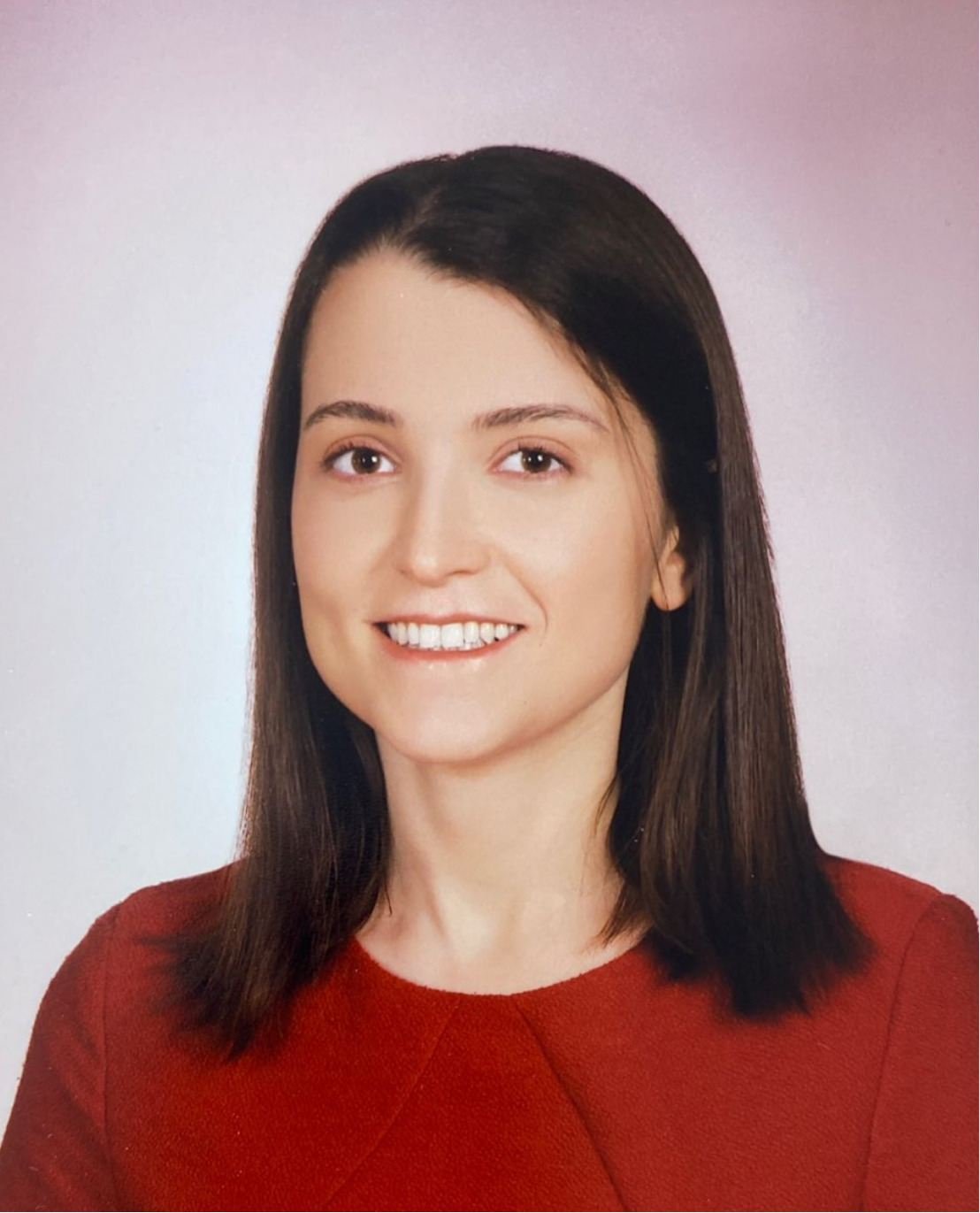}}]{Selen Gecgel Cetin}~(Graduate Student Member, IEEE) received the B.S. degree in electronics and communication engineering from Yildiz Technical University, Istanbul, Turkey, in 2016 and the M.S. degree in telecommunication engineering from Istanbul Technical University (ITU), Istanbul, in 2019, where she is currently pursuing the Ph.D. degree. From 2019 to 2021, she was a research assistant in the Department of the Turkish Air Force Academy, NDU. She is a member of ITU Wireless Communication Research Laboratory and the Poly-Grames Research Center. Her current research interests are machine learning and wireless communications.
\end{IEEEbiography}

\begin{IEEEbiography}[{\includegraphics[width=1in,height=1.25in,clip,keepaspectratio]{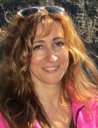}}]{Ángeles Vazquez-Castro}~(Senior Member, IEEE) received the M.Sc. and Ph.D. degrees from Vigo University, Vigo, Spain, in 1994 and 1998, respectively. Her Ph.D. degree was granted by the European Space Agency to develop land mobile satellite channel models. These models were used to develop DVB-SH standard for satellite services. From 2002 to 2004, she worked with the European Space Agency ESTEC in The Netherlands on quality of service via adaptive coding and modulation. This work is part of the guidelines of the DVB-S2 standard. Since 2004 she has been an associate professor at the Autonomous University of Barcelona, Barcelona, Spain. Her results are published in more than 130 peer-reviewed scientific papers, 12 book chapters, 1 book as editor, 2 patents and contributions to several standardization bodies. Her current research interests include classical and quantum communication networks.
\end{IEEEbiography}

\begin{IEEEbiography}[{\includegraphics[width=1in,height=1.25in,clip,keepaspectratio]{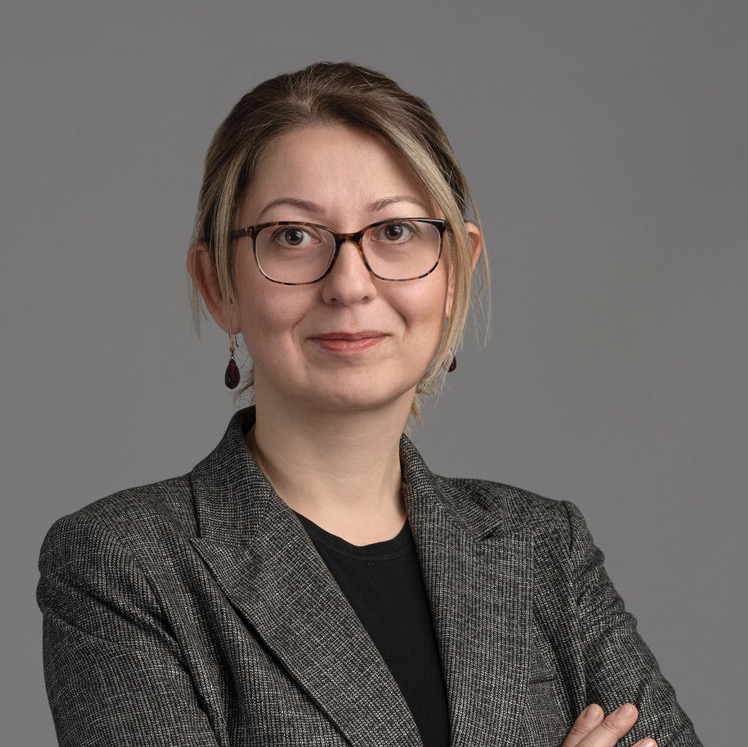}}]{Gunes Karabulut Kurt}~(Senior Member, IEEE) is a Canada Research Chair (Tier 1) in New Frontiers in Space Communications and Associate Professor at Polytechnique Montreal, Montreal, QC, Canada. She is also an adjunct research professor at Carleton University. Gunes received the B.S. degree with high honors in electronics and electrical engineering from Bogazici University, Istanbul, Turkiye, in 2000, and M.A.Sc. and Ph.D. degrees in electrical engineering from the University of Ottawa, ON, Canada, in 2002 and 2006, respectively. Between 2005 and 2010, she worked in various technology companies in Canada and Turkiye. From 2010 to 2021, she was a professor at Istanbul Technical University. Gunes is a Marie Curie Fellow and has received the Turkish Academy of Sciences Outstanding Young Scientist (TUBA-GEBIP) Award in 2019. She is serving as the secretary of IEEE Satellite and Space Communications Technical Committee, the chair of the IEEE special interest group entitled “Satellite Mega-constellations: Communications and Networking” and is also an editor for 6 different IEEE journals. She is a member of the IEEE WCNC Steering Board and a Distinguished Lecturer of the Vehicular Technology Society Class of 2022.
\end{IEEEbiography}
\end{document}